\documentclass[amsmath,amssymb,floatfix]{revtex4}
\newcommand {\eq}{\begin{equation}}
\newcommand {\qe}{\end{equation}}
\newcommand {\cen}[1]{\begin{center} #1 \end{center}}

\newcommand {\bfq}{{\bf q}}

\newcommand {\ea} {{\it et al. }}

\newcommand {\bfk}{{\bf k}}
\newcommand {\bfp}{{\bf p}}
\newcommand {\h}{\frac{1}{2}}

\newcommand {\pr}{ Phys. Rev. }
\newcommand {\np}{Nucl. Phys. }

\newcommand {\lpm}{\ell\pm}
\newcommand {\lp}{\ell+}
\newcommand {\lm}{\ell-}
\newcommand {\pth}{P3/2 }
\newcommand {\poh}{P1/2 }

\newcommand {\fsh}{F7/2 }
\usepackage{graphicx}
\usepackage{epsfig}
\begin{document}





\title{Partial-wave analysis of $K^+$ nucleon scattering}

\author{ W. R. Gibbs and R. Arceo}
\affiliation{ Department of Physics, New Mexico State University \\
 Las Cruces, New Mexico 88003, USA\\}

\begin{abstract}

\cen{\bf Abstract}

We have performed a partial-wave analysis of {\it K}$^+$-nucleon 
scattering in the momentum range from 0 to 1.5 GeV/c 
addressing the uncertainties of the results and comparing 
them with several previous analyses. It is found that the 
treatment of the reaction threshold behavior is particularly 
important. We find a {\it T}=0 scattering length which is not 
consistent with zero, as has been claimed by other analyses. 
The {\it T}=0 phase shifts for $\ell>0$ are consistent with a pure 
spin-orbit potential. Some indications for the production of 
a {\it T}=0 pentaquark with spin-parity {\it D}5/2+ are discussed.

\end{abstract}

\maketitle

\section{Introduction}

The interaction of the $K^+$ meson with the nucleon has a 
number of interesting features. Because of lack of annihilation 
of the meson anti-quark there is no three-quark intermediate 
state possible so 3-q resonances are not possible.  
This lack of 3-body states leads to a feeble interaction, among 
the smallest among the strong forces.

This fact has been useful for probing the possible changes in 
nucleon structure and/or {\it K}$^+${\it N} interaction in the nuclear 
medium \cite{siegel1,siegel2,brown,krauss,weiss,gal1,gal2,tolos}.  
Following the measurement of the ratio of total cross sections 
\cite{krauss,weiss}, a number of suggestions were made to 
explain the results, among the principal ones being a partial 
deconfinement of quarks, an interaction with exchanged mesons 
in the nucleus and a modification of the {\it K}$^+${\it N} interaction 
through exchanged mesons. In order to understand which of 
these possible scenarios might be the right one, a detailed 
comparison with experiment is in order. One element in making 
calculations of the multiple scattering corrections which are 
crucial for this comparison is the availability of reliable 
phase shifts.

Of equal importance is the use of the {\it KN} system 
to test fundamental theories of hadronic interaction. There 
have been a number of theoretical studies of the {\it K}$^+${\it N} 
system using a variety of 
approaches \cite{barnes,buttgen,bender,campbell,veit,cohen}. 
The phase shifts from {\it K}$^+${\it N} scattering provide the principal 
data with which they compare.

There have been very little data taken (but see 
\cite{krauss,weiss}) in an ordinary sense since the most 
recent partial-wave analyses 
\cite{martinoades,hash,martinrome,martin,arndt2,arndt1,hyslop,corden, 
martin78,martinpl,nakajima,kato,cutkosky,watts}. The 
suggestion of the existence of a possible pentaquark 
resonance (see Ref. \cite{penta} for a list of experiments 
claiming to see it, some of which have been withdrawn) might 
change the point of view of any analysis. The existence of 
the pentaquark is now considered doubtful by many.

A partial-wave analysis is particularly interesting in the 
region of the reaction threshold. The production of pions in 
the collision of two nucleons is often calculated with the 
``re-scattering'' diagram which describes the formation of 
the $\Delta^{33}$ resonance by the exchange of one pion. In 
the case of $K^+$ scattering this simple re-scattering 
process does not exist since the $K^+$ cannot emit a single 
pion, and other mechanisms must be considered. The excitation 
by exchange of a $\rho$ meson might be a natural process to 
consider, at least from the point of view of the exchange of 
heavy mesons.  Additional information on this process can be 
obtained from a phase shift analysis which indicates 
which partial waves are participating in the production.

Section II describes our fitting method, Section III gives
a description of the  T=1 analysis, Section IV discusses the
results of the T=1 analysis, and Section V gives the method
for the T=0 analysis. Section VI discusses the results of the
T=0 analysis and Section VII gives an overall summary and
discussion of the results.

\section{Fitting Procedure}

In the heart of any amplitude analysis lies the search for the 
(or a) minimum in a $\chi^2$ (or similar) measure for the best 
fit to the data. It is normally assumed that the numerical 
procedure for this minimization is straightforward and does not 
pose any problems, but that may not be the case.  The technique 
that we use is very pedestrian but seemingly very sure.  The 
$\chi^2$ is minimized on each parameter in turn sweeping through 
a significant number of them (88 for the {\it T}=1 case and 61 for the 
{\it T}=0 case including data normalizations). To minimize $\chi^2$, 
each parameter is stepped by a fixed interval until the value of 
$\chi^2$ increases.  At this point in the search three values of 
$\chi^2$ are known at three values of the parameter. A parabola 
is then passed through the three points and the position of the 
minimum is predicted.  This procedure, in some form, is common 
to most methods of minimization although several methods treat 
the full parameter space as a vector. There is a difficulty that 
arises due to the fact that $\chi^2$ as a function of the 
parameter is very often not a perfect parabola. This means that 
the prediction of the minimum position (and hence $\chi^2$ 
value) is not the true minimum in this region and the value of 
$\chi^2$ predicted may exceed the one at the central (lowest) 
point. Since the middle point is very often the value obtained 
in the search on the previous parameter in the sequence, if the 
predicted value is always accepted, the ''current minimum'' 
$\chi^2$ will increase in some cases. In our algorithm we test 
the predicted value of $\chi^2$ against the central value and if 
it is greater, the central (previous best) value is used 
instead.

This incorrect prediction is not a rare occurrence. We observed 
that it happened about 4\% of the time when the search was far 
from the minimum and up to 40\%-50\% of the time when it was 
close to the minimum. If one decreases the step size, the 
deviation from a parabolic shape can be lessened, but there is a 
limit to this process since the difference between values of 
$\chi^2$ needed to predict the new minimum becomes small 
compared with their values and with finite precision another 
source of error becomes important, even with double precision 
(which we use). Hence, there is an inherent limit to how well 
the minimum can be found.

This limit depends on the details of the method used, of course, 
but also on the parameterization of the phase shifts. Different 
parameterizations will have different functions to replace the 
parabola, or perhaps more practically, will have a different 
importance of third order terms. 

This inability to find perfectly the minimum (or minima) 
translates to a dependence of the final result on the starting 
values. Tests of the dependence on starting point were made by 
perturbing some of the parameters (typically two or three) until 
the $\chi^2$ was very large.  For $\chi^2$ of the order of 
100,000 the search usually was unable to find a sensible minimum 
but for $\chi^2$ in the range 10,000 to 20,000 a minimum was 
found near, but not identical, to the principal fit.  A large 
number of sweeps through the parameters was needed (several tens 
of thousands). The results of these tests are given in the 
discussions of the {\it T}=1 and {\it T}=0 analyses.

A $\chi^2$ corresponding to each normalization was included in the 
total $\chi^2$ of the fit by adding

\eq
\left(\frac{N-1}{\Delta N}\right)^2\label{chinorm}
\qe
for each normalizing parameter, N, to the $\chi^2$ coming from 
the individual data points. The value of 0.03 was chosen for 
$\Delta N$ since it is a typical value for normalization errors.

The philosophy behind this procedure is that the normalization 
should be treated as an independent data point. In a typical 
model experiment the number of counts is registered in a set of 
counters and then those counts are multiplied by a normalization 
factor determined by the effective number of target particles, 
the beam flux etc. The errors in this normalization factor will 
be common to the cross sections for every counter. While this 
would appear to be the right statistical procedure, some care is 
needed in the interpretation of the results. Even if a 
significant discrepancy occurs in the normalization, it can be 
hidden if one simply looks at the total $\chi^2$ per data point 
in the case that there are a large number of points which 
accompany a single normalization. The large number of points 
dilute the total $\chi^2$ so that it is only slightly larger 
than the number of degrees of freedom.  A distinction based 
solely on total $\chi^2$ requires the separation of a value of 
the reduced $\chi^2$ closer and closer to unity from unity. 
Practical difficulties arise since one can assume that the 
experimental errors have been estimated correctly only up to a 
certain level (experience would suggest something of the order 
of 5\%). Hence the square of the errors in the denominator of 
the definition of $\chi^2$ leads to a sufficient error in its 
value that this method is impractical.  This does not mean that 
a problem with the normalizations cannot be found, however. It 
simply means that one must examine the normalizations separately 
to verify that the the $\chi^2$ from the normalizations alone is 
not excessive.

\section{$K^+$ proton ({\it T}=1) analysis}

The {\it T}=1 data base used consists of 1880 data points including 
1501 angular distribution points, 91 total cross section points, 
265 polarization points and 23 reaction cross section points. 
Much of these data are available from compilations 
\cite{compilation,vpidb,pdg} but we try to give references to 
the original data as much as possible 
\cite{ehrlich,bugg68,caldwell70,cool70,abrams70,bowen70,loken72,bowen72, 
barnett73,burnstein74,abe74,patton75,adams,cook,blandr,blanda,bettini,filippas, 
giakpp,berthon,fisk}.

Of these points, four were removed because they were outliers. 
These were: two points from Ref. \cite{bowen70} at 0.686 and 0.717 
GeV/c, and two from Ref. \cite{bugg68} at 0.713 and 1.029 GeV/c. 
The error on Ref.  \cite{ehrlich} polarization at 1.330 GeV/c and 
$\cos\theta=0.242$ was doubled. The errors were doubled on the 
polarization set at 1.430 GeV/c \cite{ehrlich} since the variation 
from point to point is greater than the errors quoted.

\begin{figure}[htb]
\epsfig{file=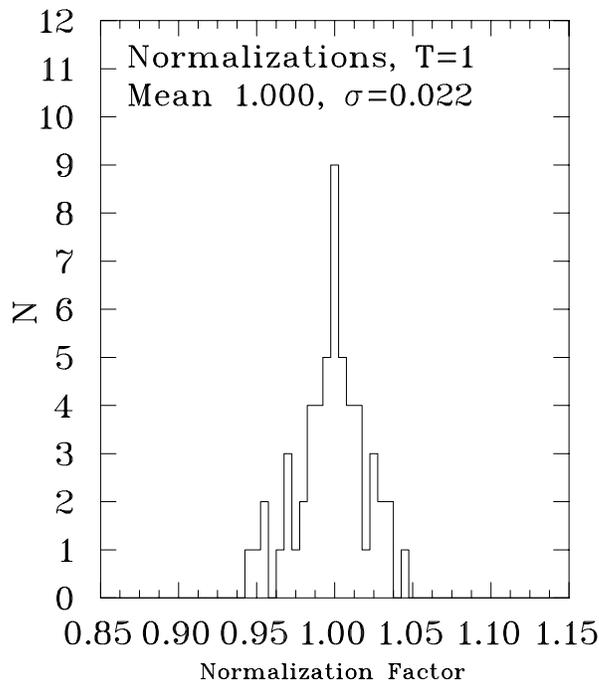,height=3.5in}
\caption{Distribution of the normalization factors for isospin 1.} 
\label{norms1} \end{figure}

\begin{table}
$$ \begin{array}{|c|c|r|r|c|c|}
\hline
{\rm LJ}&q_R\ {\rm (GeV/c)}&\gamma_1\ \ \ &\gamma_2\ \ \ 
&\gamma_3&{\rm Equation}\\
 \hline
 S1/2 &    0.4134 & 0.0858  & 0.5030&&\ref{etas}\\
 \hline
 P1/2 &  0.3858  & 0.1811 & -0.0163&&\ref{etas}\\
 \hline
 P3/2 &    0.4809 &  1.0794 &  2.1643 & 
-3.9024&\ref{etas}\\
 \hline
 D3/2 & 0.6758  & 2.1264&&&\ref{etad32}\\
 \hline
 D5/2 & 0.4304 &0.3604 & -0.2723&&\ref{etas}\\
 \hline
 F5/2  & 0.5286  & 0.1778&&&\ref{etas}\\
 \hline
\end{array}
$$
\caption{Parameters for the representation of the inelasticity,
$\eta_{\ell,j}$ for {\it T}=1 using the form of Eq. \ref{etas} except
for the {\it D}3/2 wave where Eq. \ref{etad32} is used.}\label{parseta1}
\end{table}

\begin{table}
$$ \begin{array}{|r|r|r|r|r|c|}
\hline
LJ&a\ {\rm (GeV/c)}^{-(2\ell+1)}&b_{1,\lpm}\ {\rm 
(GeV/c)}^{-2}&b_{2,\lpm}\ 
{\rm (GeV/c)}^{-4}&b_{3,\lpm}\ 
{\rm (GeV/c)}^{-6}&{\rm Equation}\\
 \hline
 S1/2 &    -1.562 & -1.108 &  0.217&&\ref{spwaves}\\
 \hline
 P1/2 &  -12.002  &31.139&&&\ref{spwaves}\\
 \hline
 P3/2 &   13.357  & 126.676 & -666.951 & 1276.123
&\ref{spwaves}\\
 \hline \hline
LJ&c_{\pm}\ {\rm (GeV/c)}^{-5}&d_{\pm}\ {\rm (GeV/c)}^{-7}&e_{\pm}\ 
{\rm (GeV/c)}^{-9}&&{\rm Equation}\\
\hline
 D3/2 & -2.984 &  4.119&&&\ref{dwaves}\\
 \hline
 D5/2 & -1.702 &  6.571 & -7.462&&\ref{dwaves}\\
 \hline \hline
LJ&f_{\pm}\ {\rm (GeV/c)}^{-7}&&&&{\rm Equation}\\
 \hline
 F5/2  & -0.415&&&&\ref{fwaves}\\
 \hline
 F7/2 & 0.089&&&&\ref{fwaves}\\
\hline
\end{array}
$$
\caption{Parameters for the representation of the phase shifts
for {\it T}=1 using the form of Eq. \ref{spwaves}
for the {\it S} and {\it P} waves, Eq. \ref{dwaves} for the {\it D}
 waves and Eq. \ref{fwaves} for the {\it F} waves.}\label{parsdel1}
\end{table}

Partial waves through \fsh were considered. The {\it G} and {\it H}
waves were not needed to get a good fit. We write the partial-wave 
amplitudes as

\eq
F_{\lpm}=\frac{(S_{\lpm}-1)e^{2i\sigma_{\ell}}}{2i};\ \ 
S_{\lpm}=\eta_{\lpm}e^{2i\delta_{\lpm}}
\qe
where the sign $\pm$ corresponds to $j=\ell\pm\h$ and 
$\sigma_{\ell}$ is the Coulomb phase shift. No ``inner''
Coulomb corrections were made, the Coulomb effect being very
small for the energies considered here.

The differential cross section, polarization and total cross 
section are expressed in terms of the amplitudes in the standard 
manner (see e.g. Hashimoto \cite{hash}) and will not be repeated 
here. The reaction cross section (essentially pion production) can 
be written as
\eq 
\sigma_R=\frac{10\pi}{q_f^2}\sum_{\ell}[\ell(1-\eta^2_{\lm})+ 
(\ell+1)(1-\eta^2_{\lp})] 
\qe 
so is independent of the phase shifts and can be particularly 
important in determining the $\eta_{\lpm}$. Here, and in what 
follows, $q$ denotes the momentum in the center of mass (in 
GeV/c unless otherwise noted) and $q_f$ is $q/\hbar c$, i.e., in 
femtometers.

We have taken for the {\it S}- and {\it P}-wave phase shifts the form
\eq
\delta_{\lpm}=\tan^{-1}\left[\frac{a_{\lpm}q^{2\ell+1}}
{1+\sum_{i}b_{i,\lpm}q^{2i} }\right].\label{spwaves}
\qe
The index, $j$ runs from 1 to 2 for the S wave and 1 to 3 for the \pth wave.
There is only one term in the sum for the \poh wave. 

For the {\it D} waves a simple polynomial is used
\eq
\delta_{2\pm}=q^5(c_{\pm}+d_{\pm}q^2+e_{\pm}q^4)\label{dwaves}
\qe
and only the lowest order was used for the {\it F} waves
\eq
\delta_{3\pm}=q^7f_{\pm} \label{fwaves}
\qe

The $\eta_{\lpm}$ were parameterized allowing a different 
threshold, $q_R$, for each partial wave. Below the threshold 
the values were unity and above threshold the form

\eq
\eta=1-\sum_{i=1}^n\gamma_i\left(\frac{q}{q_R}-1\right)^i\label{etas}
\qe
was used except for the D3/2 partial wave which was represented by
\eq
\eta=\cos \left[\sum_{i=1}^n\gamma_i\left(\frac{q}{q_R}-1\right)^i\right].
\label{etad32}
\qe
The highest power of $j$ varied with partial wave, the greatest 
being 3 in the case of the \pth wave. For the \fsh wave $\eta$ was 
taken as unity. The values of the parameters for our best fit are 
given in Tables \ref{parseta1} and \ref{parsdel1}.

The normalizations of 55 of the data sets was varied in the 
fitting procedure and their distribution is given in Fig. 
\ref{norms1}. The standard deviation of the normalizations was 
2.2\%, less than the 3\% used for $\Delta$N in Eq. 
\ref{chinorm}. The mean normalization was 1.0004 and only two 
normalizations gave an adjustment greater than 5\%. The total 
$\chi^2$ is 2051 for a $\chi^2$ per data point of 1.09.

\begin{figure}[htb]
\hspace*{-.1in}\epsfig{file=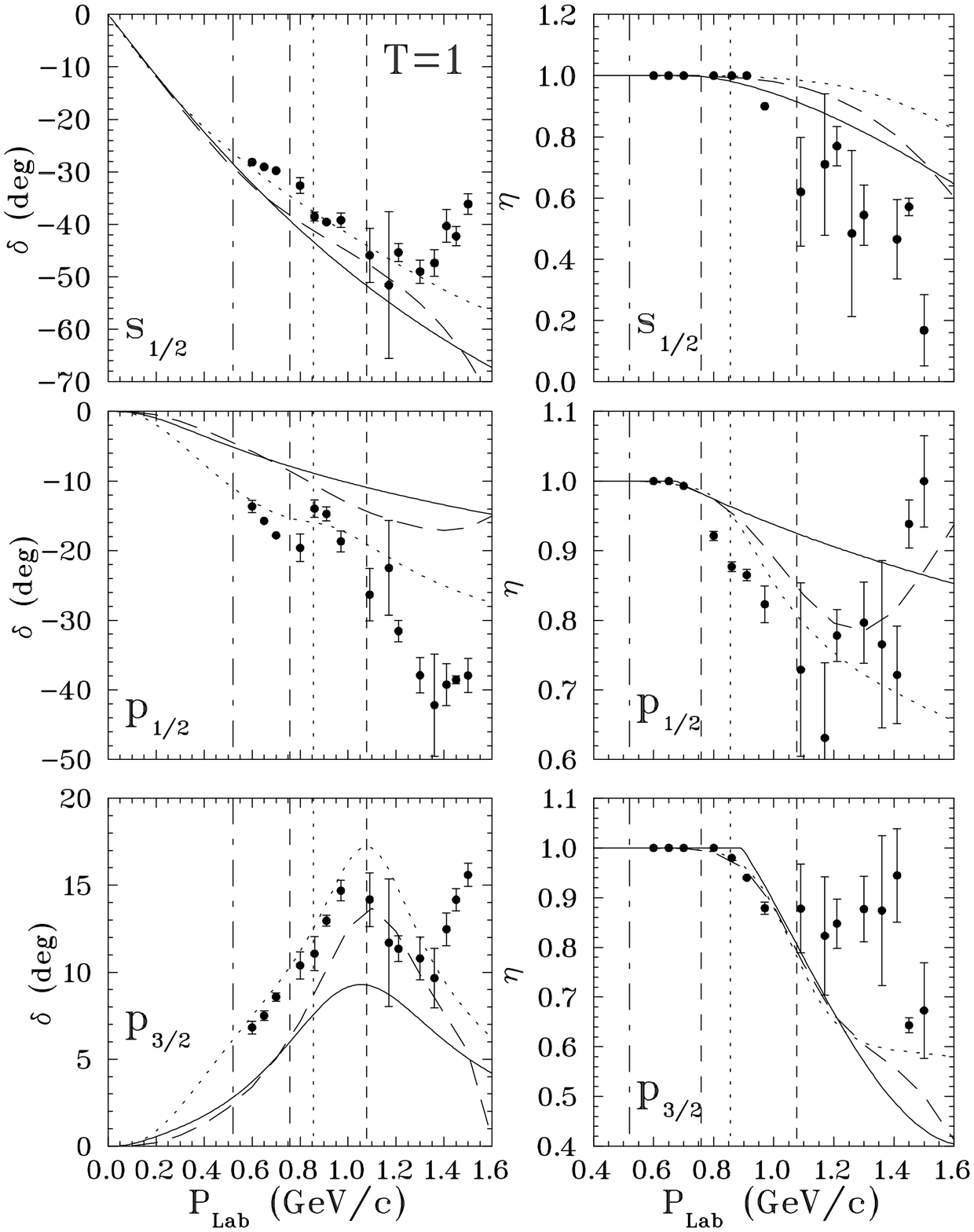,height=3.9in}
\epsfig{file=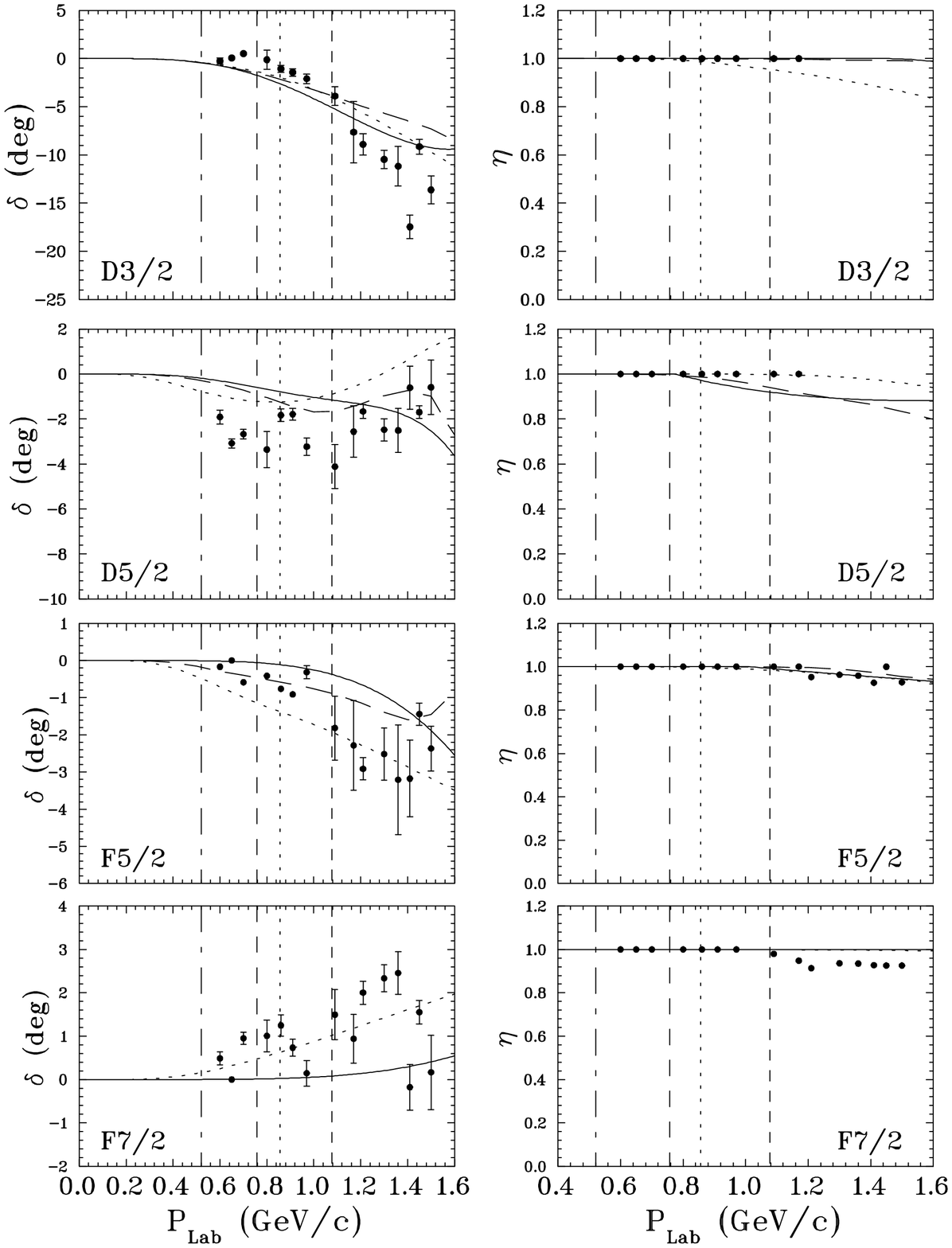,height=3.9in}
\caption{T=1 phase shifts and $\eta$'s obtained in the present 
work (solid line) compared with VPI (dotted line), Hashimoto 
(dots) and Martin (dashed line). The vertical lines show the 
thresholds; dash-dot: pion production threshold, long-dashed:  
threshold for production of a theta particle and a 
pion, dotted: threshold for production of a delta and 
a $K^+$, and short-dashed: production by a K$^*$(892) nucleon.} 
\label{phases1} \end{figure}

\begin{figure}[htb]
\hspace*{-.1in}\epsfig{file=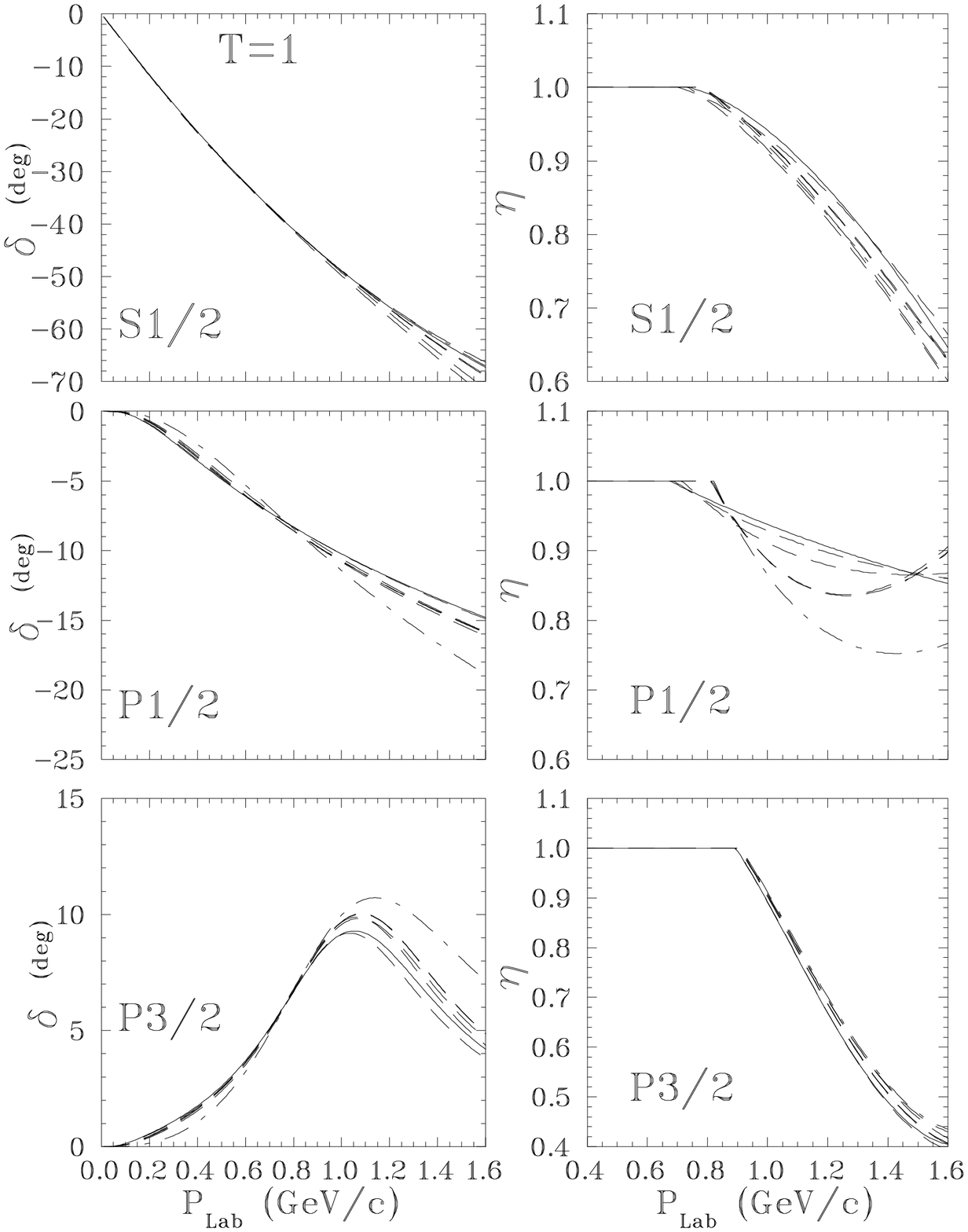,height=3.9in}
\epsfig{file=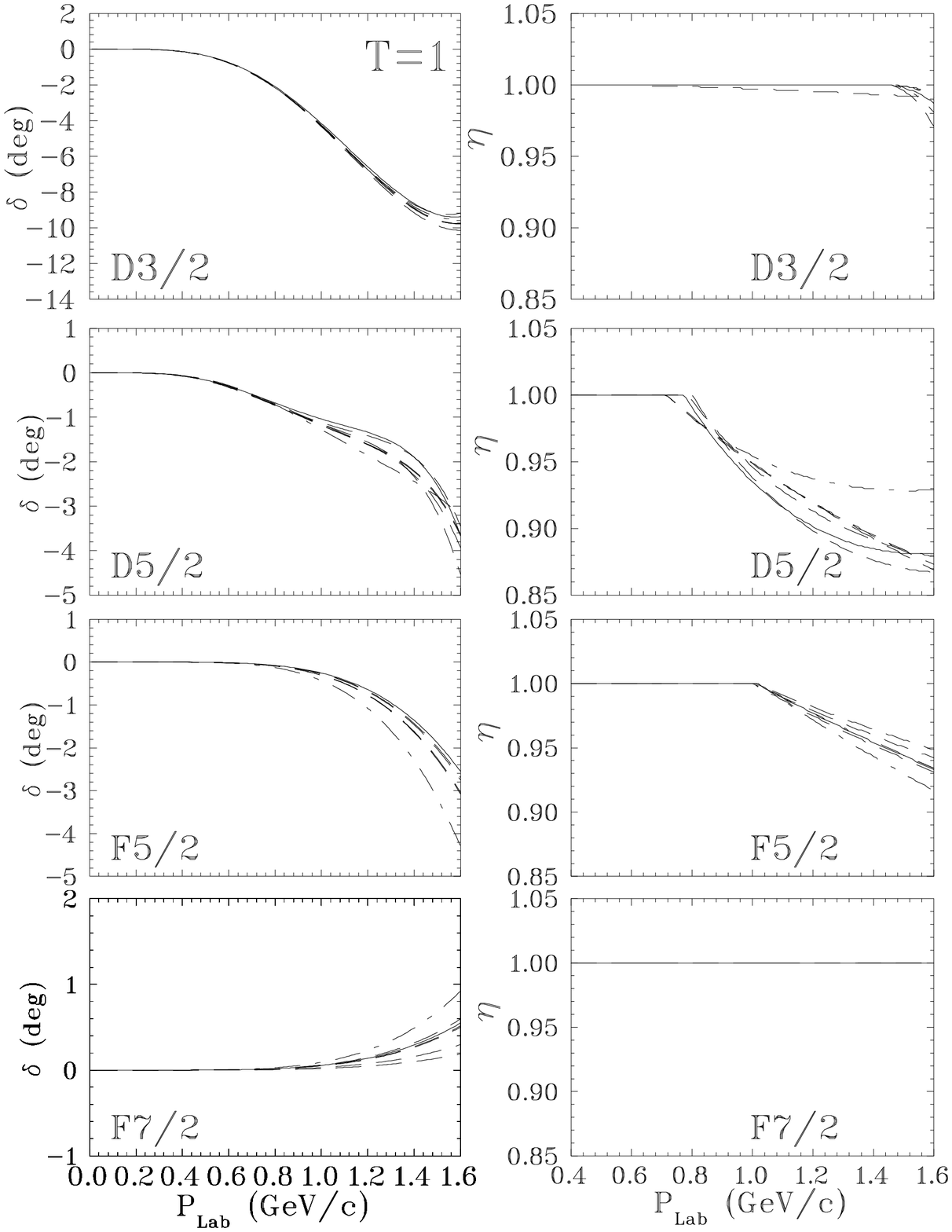,height=3.9in}
\caption{T=1 phase shifts and $\eta$'s obtained in the present 
work (solid line) compared with the variations. The solid line 
is the principal fit, the dashed lines correspond to fits a, b, 
c, e, and f, and the dash-dot line to the poorer fit d. The designations
of the cases correspond to those in Table \ref{minima1}.} 
\label{compare1} \end{figure}

The phase shifts for the best fit are shown in Fig. \ref{phases1} 
compared with three other analyses discussed below. At low energy 
(below threshold) the results are very similar to those of Martin 
\cite{martin}.

Aside from the main fit, we performed tests for the degree of 
uniqueness of the minimum. Values of the parameters were altered in 
varying degrees such that the $\chi^2$ values were very large 
(generally of the order of 10,000) and the minimum search was 
restarted. This was done 6 times and the values of $\chi^2$ at the 
minima found are given in table \ref{minima1}. The values of 
$\chi^2$ were about 9 above the best fit except for case d where 
$\chi^2$ was about 20 above the best fit.

\begin{figure}[htbp]
\epsfig{file=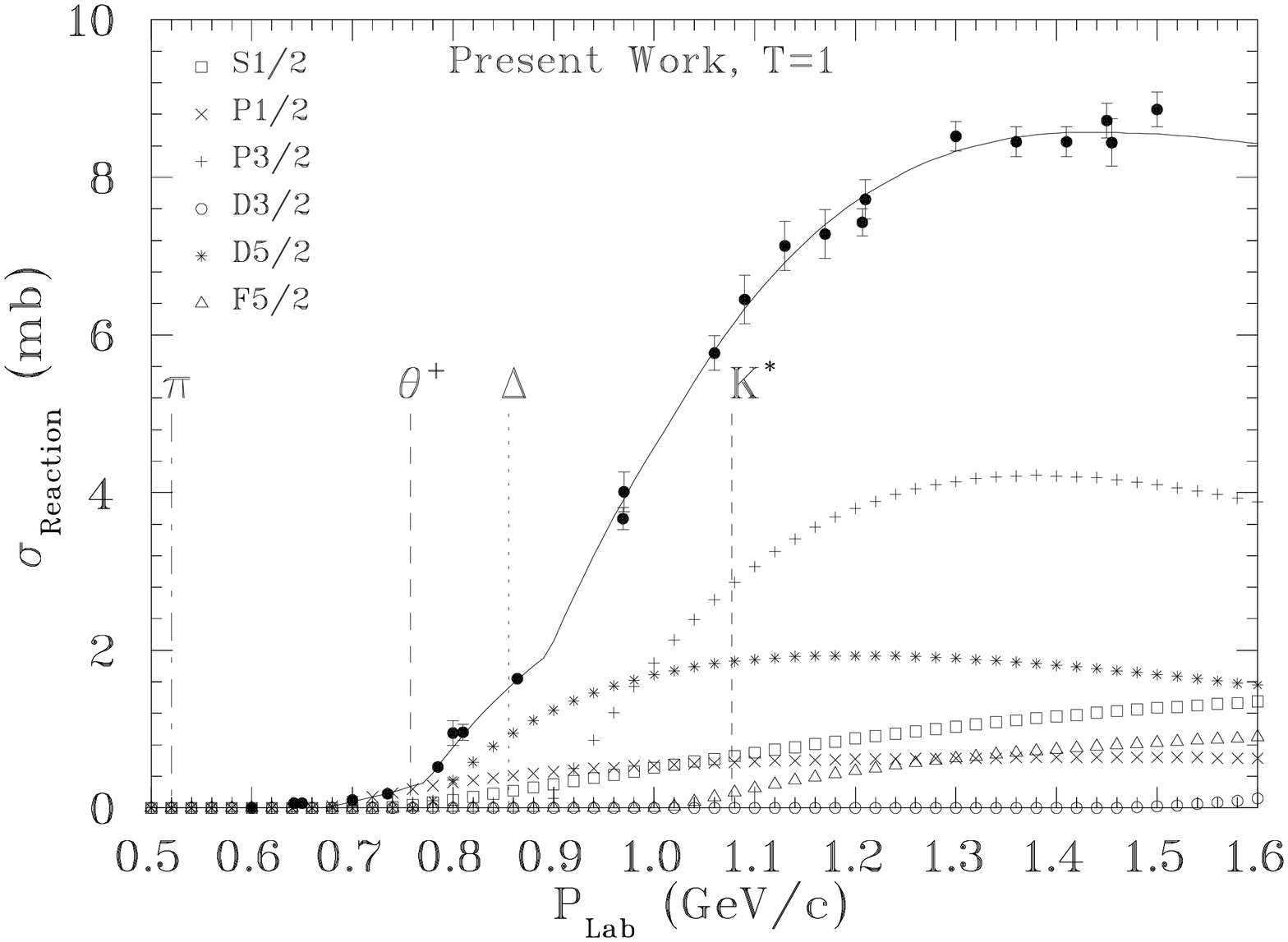,angle=90,height=4in}
\epsfig{file=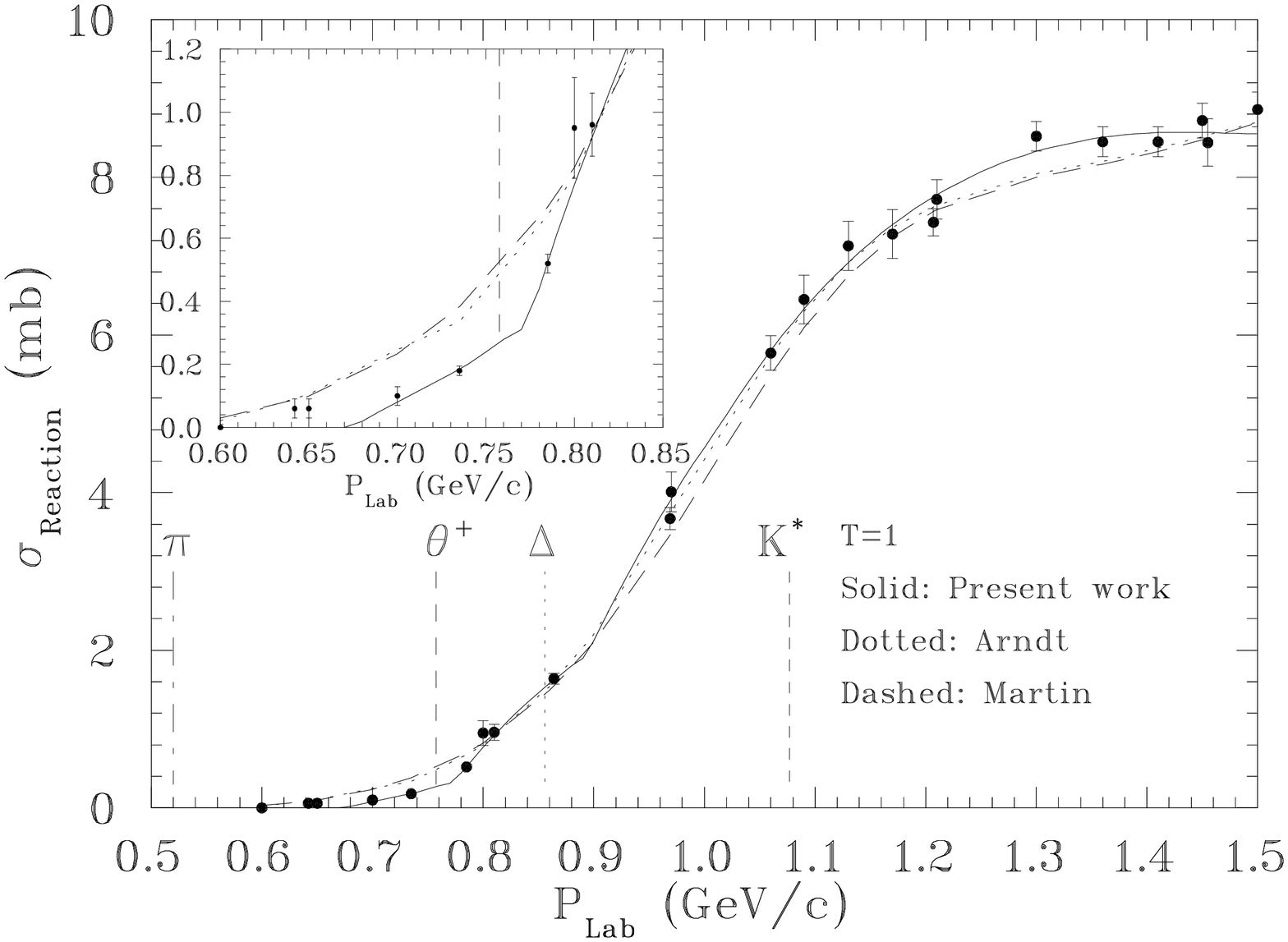,angle=90,height=4in}
\caption{Top: Contribution of the various partial waves to the 
reaction cross section. Bottom: Comparison of the reaction cross 
section from the present work with that of the VPI group and 
Martin.} \label{reaction} \end{figure}

The fits found at these other minima are very similar to the 
original fit. The comparison is shown in Fig. \ref{compare1}. 
There it is seen that the fits are nearly identical below 
threshold (with the possible exception of case d) but above 
threshold there is a noticeable variation. In principle one can 
eliminate all fits but the principal solution on the grounds 
that the difference in $\chi^2$ is considerably larger than 
unity. In practice, however, since one can never be sure that 
the ``true'' minimum has been found, we consider the variation 
among the fits (excluding fit d) as a conservative estimate of 
the error in the determination of the phase shifts and 
inelasticities.

\begin{table}
$$ \begin{array}{|c|c|c|c|c|c|}
\hline
{\rm Case}&\chi^2&\chi^2/N&{\rm Scatt.\ Len.\ (fm)}&{\rm P1/2\ (fm}^3{\rm )}
&{\rm P3/2\ (fm}^3{\rm )}\\
 \hline
{\rm Best\ Fit}  &  2031.05 & 1.080 &-0.308&-0.092&0.103\\ 
 \hline
a   & 2042.47  & 1.086 &-0.311&-0.055&0.046\\ 
 \hline
b   & 2039.94 & 1.085 &-0.310&-0.067&0.058 \\ 
 \hline
c   & 2040.00  & 1.085 & -0.310&-0.066&0.057\\ 
 \hline
d   & 2051.98  & 1.091& -0.313&-0.029&0.009\\ 
 \hline
e  &  2039.67  & 1.085 & -0.308&-0.084&0.094 \\ 
 \hline
f  &  2039.02  & 1.085& -0.311&-0.060&0.051\\ 
 \hline
\end{array}
$$
\caption{Values of $\chi^2$ for the best fit and the 
variations made in the present work for T=1.
The columns labeled P1/2 and P3/2 contain 
scattering volumes.\label{minima1}} 
\end{table}

\subsection{Other Work for T=1}

Leaving aside many of the earlier analyses 
\cite{martinoades,martinrome,corden, 
martin78,martinpl,nakajima,kato,cutkosky,watts} there are three 
modern analyses with which we compare.

\subsubsection{Analysis of Hashimoto}

Hashimoto \cite{hash} performed an energy-independent analysis 
for momenta from 0.6 to 1.5 GeV/c. Such an analysis has the 
advantage that no theoretical prejudices are inserted in the 
parameterization of the energy dependence but the disadvantage 
that S-matrix elements obtained at one energy do not share 
information from nearby energies which can result in large 
fluctuations from one energy to another. Observables sometimes 
have to be grouped to have enough data at a given energy. When 
amplitudes are slowly changing this does not lead to problems 
but at the threshold for pion production, where some of the 
amplitudes change rapidly, it can. Structure was seen in this 
analysis in the \pth and other partial waves.

\subsubsection{Analysis of the VPI Group}

The VPI group \cite{arndt2,arndt1,hyslop} has published three 
analyses. In these fits the S-matrix elements for each partial wave 
were parameterized in the form of a K-matrix with one elastic and 
one inelastic channel. Structure is seen in these fits in the \pth 
wave very similar to that of Hashimoto. This led the group to 
suggest that there was a resonance in this wave and two analyses 
give the mass pole at 1.780 and 1.796 GeV/c$^2$, corresponding 
to beam momenta of 0.971 and 1.005 GeV/c. 

\begin{figure}[htb]
\epsfig{file=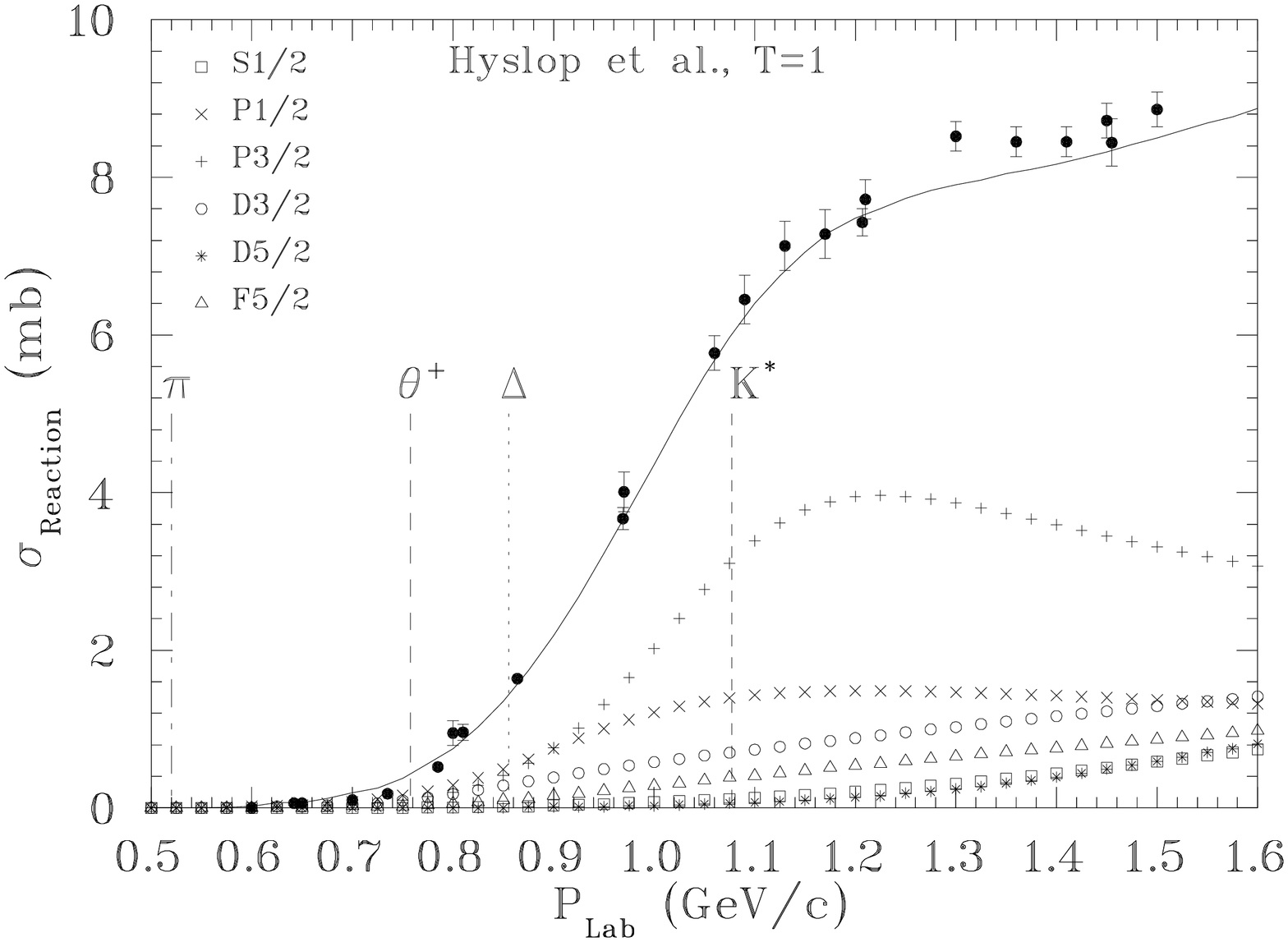,angle=90,height=4.5in}
\epsfig{file=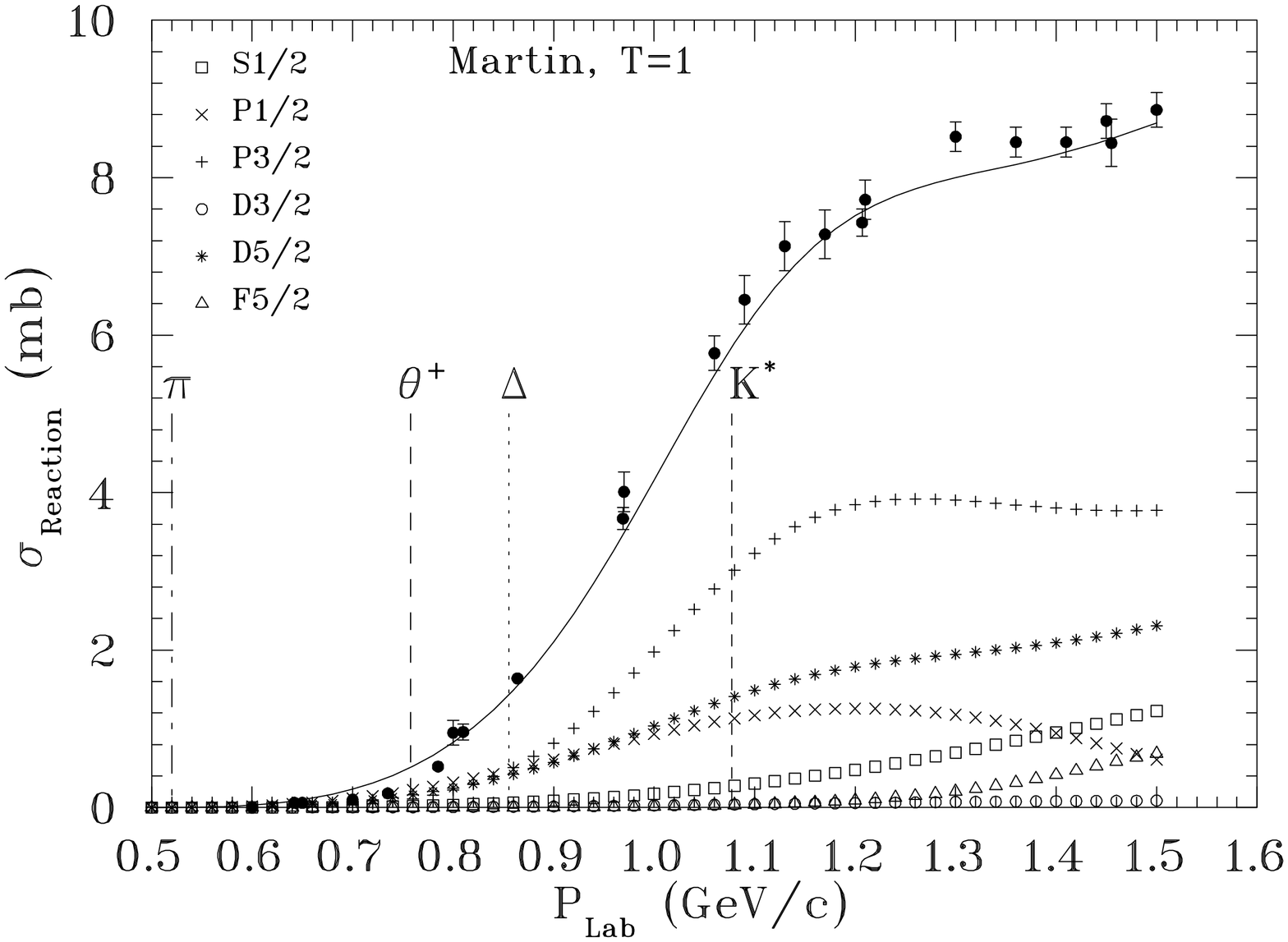,angle=90,height=4.5in}
\caption{Partitioning of the {\it T}=1 reaction cross sections among 
partial waves for Hyslop \ea \cite{hyslop} and Martin 
\cite{martin}.}
\label{areact} \end{figure}

The onset of one-pion inelasticity was assumed to come about by 
the production of intermediate particles which then decayed to 
the $KN\pi$ system. Such a model is made very plausible by the 
experimental fact that pion production does not start at its 
threshold but some 200 MeV/c higher in momentum (see Figs. 
\ref{reaction} and \ref{areact}). Since the $\Delta$ threshold 
is 340 MeV/c above pion threshold and the delta has considerable 
width, so that the production can start below that, it was the 
prime candidate.  The {\it K}$^*(892)$ was considered as well but it 
has a higher threshold.

The VPI work \cite{arndt1,hyslop} allowed renormalization in 
much the same way as described above in the present work. 
However, unlike the present case, substantial renormalization 
occurred with a number of the data sets being renormalized by
more than 5\%.

We inserted the VPI solution (including the G and H waves) into 
our program to compare with the data base used here to find a 
$\chi^2$ of 2810 for the 1880 points (a ratio of 1.49), which 
implies a fit very similar in quality to the original fit which 
was 1.73 or 1.25 depending on the case. Two points in the 
threshold of the reaction cross section (at 0.735 and 0.785 
GeV/c) contribute 270 to the $\chi^2$. If we remove these two 
points the ratio drops to 1.35. One cannot expect a closer 
agreement since their parameters were not fit to our data base 
and the phase shifts taken from the paper \cite{hyslop} (or the 
web site) are quoted with a limited accuracy.

\begin{figure}[htb]
\epsfig{file=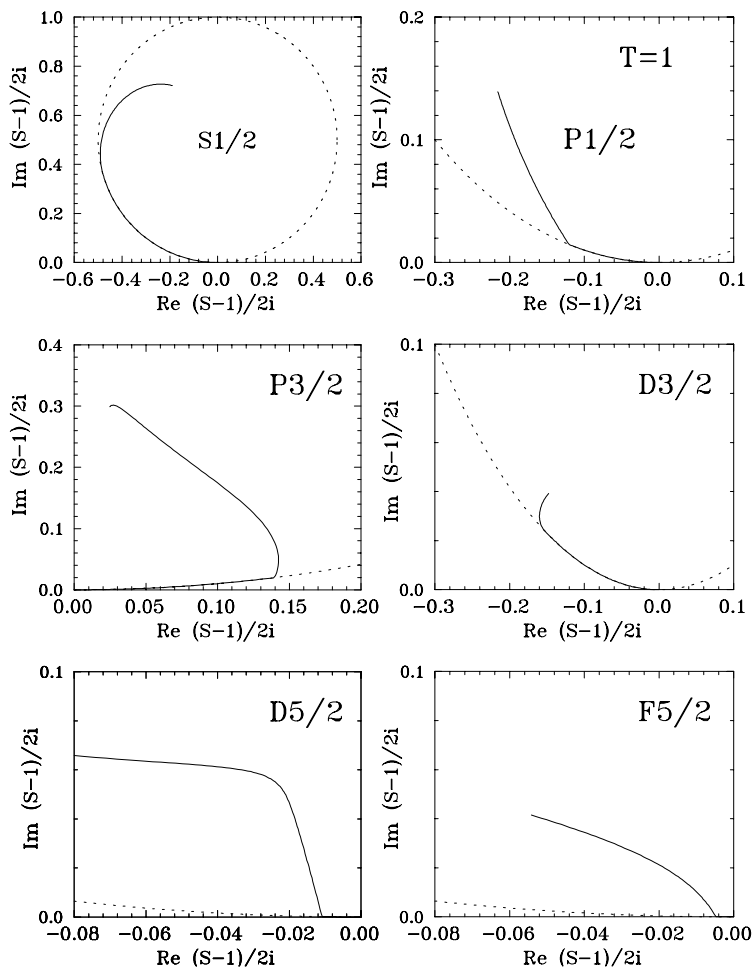,height=5.in}
\caption{Argand plot of the {\it T}=1 partial waves.}
\label{argand1} \end{figure}

\begin{figure}[htb]
\epsfig{file=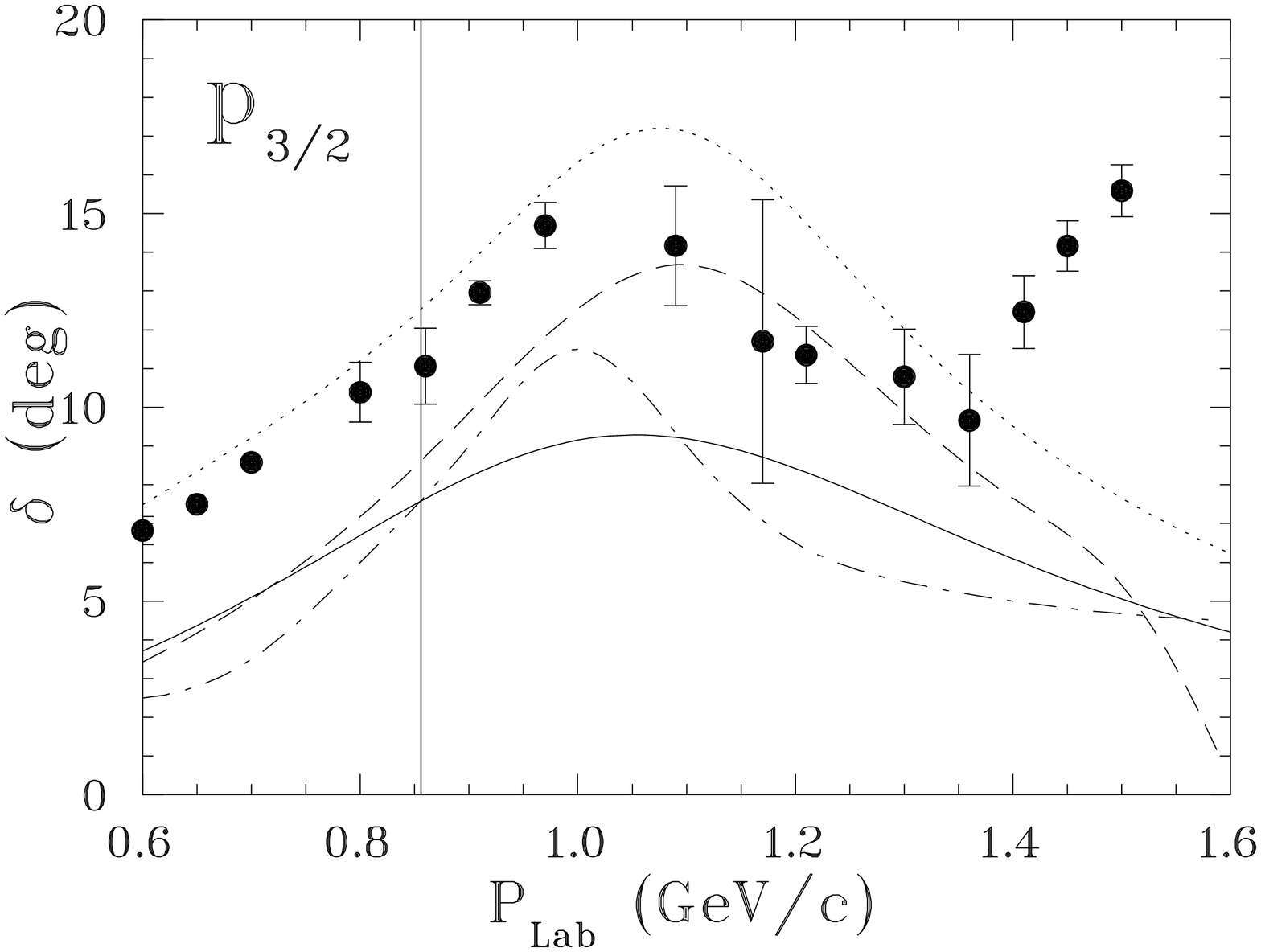,angle=90,height=3.in}
\caption{{\it T}=1 P3/2 phase shift in the region just above threshold.
The vertical line marks the $\Delta$ threshold. The chain-dash curve,
taken from the paper by Wyborny \ea \cite{wyborny} and is 
the result of a calculation of the final-state interaction between 
the {\it K}$^+$ and the $\Delta$. The other curves have the same meaning 
as in Fig. \ref{phases1}.}\label{wyborny}
\end{figure}

\subsubsection{Analysis of Martin}

Martin performed an analysis in which the partial-wave amplitudes 
were parameterized directly. The result is equivalent to the 
partial-wave expansion in terms of $\delta_{\lpm}$ and 
$\eta_{\lpm}$ discussed above and the correspondence is easily 
made.

Suppressing the partial-wave index, the Martin amplitudes have 
the 
form
\eq
F=\frac{C(q)}{A(q)+i[1+\theta(q-q_0)R^2(q)]C(q)}
\qe
\eq
C_{\lpm}=\frac{q^{2\ell+1}}{1+(q/q_0)^{2\ell+1}}.
\qe
$A(q)$ was approximated by a polynomial in $q/q_m$ and 
$R(q)$ a polynomial in $\left(\frac{q-q_0}{q_m-q_0}\right)$ 
where $q_0$ is the reaction (pion production) 
threshold (0.3106 GeV/c in the center of mass) and $q_m$ is
the maximum C. M. momentum considered in the fit.  Thus, the 
inelasticity  (expressed in terms of $R$) has a form similar to 
that used in the present work but only a single threshold 
($q_0$) was used for all partial waves.

In his fit Martin used {\it G} and {\it H} waves which were estimated by 
means of a dispersion relation calculation \cite{alcock}.  He 
did not give these values and they are no longer available so 
we were forced to set them to zero.  These waves are only 
important at the higher energies so did not affect the 
calculations done for the nuclear scattering 
\cite{siegel1,siegel2}.  However, for momenta above $\approx$ 
1.2 GeV/c the Kaon-nucleon amplitude can not be accurately 
obtained from Martin's phase shifts alone.

Martin listed several re-normalizations chosen by hand. The 
average value of these re-normalizations was 0.98 and the 
standard deviation was 0.067. Since these normalizations were 
changed by hand, no $\chi^2$ was accorded to them by Martin. 
With a fixed error of 0.03 they would have contributed 64.7 to 
the $\chi^2$.

\begin{figure}[htb]
\hspace*{-.5in}
\epsfig{file=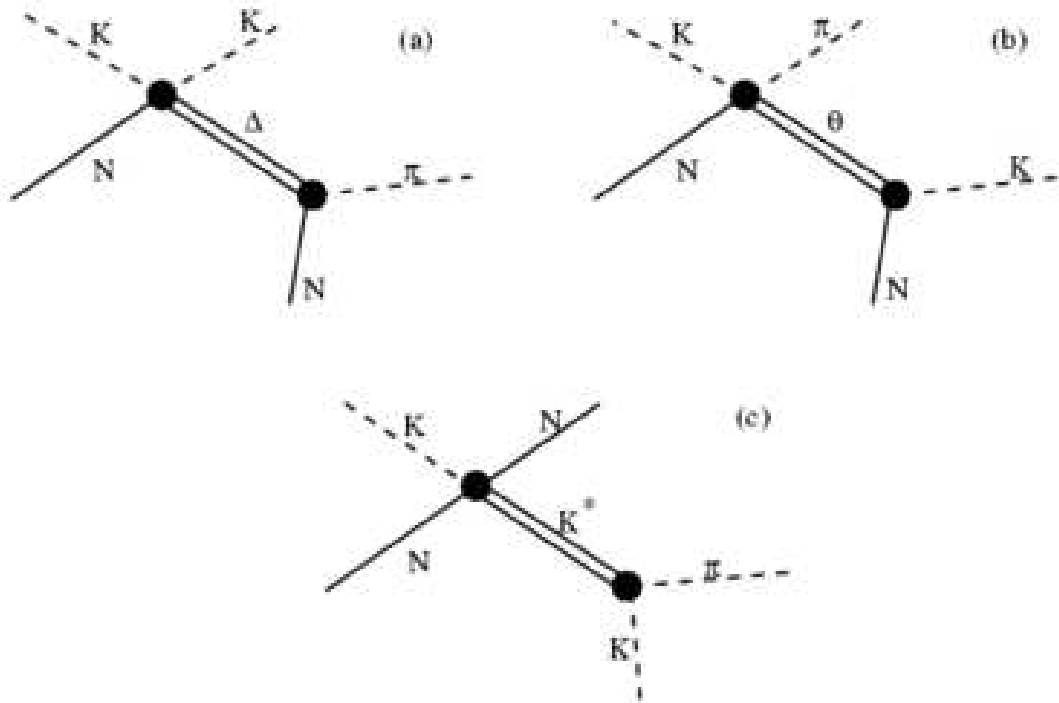,height=5in}
\caption{Diagrams for pion production. Compound objects are indicated
by the double line.} 
\label{diap1} \end{figure}

The Martin solution was also compared with our data base and we 
found a $\chi^2$ of 3023 for a ratio of 1.61. The same two 
reaction cross section points which contributed large $\chi^2$ 
in the case of the VPI group give a contribution to the $\chi^2$ 
of 446. If we remove these two points, the ratio becomes 1.37. 
Since the normalizations for the data were refit the test is not 
completely valid. The major discrepancy occurs for large beam 
momenta where the lack of {\it G} and {\it H} partial waves used in the 
original fit is most important. Removing the two reaction cross 
section points and all data above 1.2 GeV/c, we find a $\chi^2$ 
of 1372 for the 1132 points of the reduced data set or a ratio 
of 1.21 so that the quality of the fit approaches that quoted in 
the original paper which was 1.08.

\section{Discussion for {\it T}=1}

The values of the phase shifts and inelasticities for the 
obtained partial waves are given, compared with other analyses, 
in Fig. \ref{phases1}. The behavior of the amplitudes is also 
given in an Argand diagram in Fig. \ref{argand1}. We note that 
there is no counter clockwise behavior in the P3/2 partial wave 
as was observed in Nakajima \ea \cite{nakajima} or Arndt \ea 
\cite{arndt1,arndt2}.

\subsection{Reaction Cross Section}

The partitioning of the reaction cross section among the partial 
waves is given in Fig. \ref{reaction}. The lower part of Fig. 
\ref{reaction} gives the comparison of the reaction cross 
sections calculated with those from Arndt \ea \cite{arndt2} and 
Martin \cite{martin}. One notices that just above the threshold 
for pion production the reaction cross section remains very 
small for about 200 MeV/c. A similar thing is seen in 
nucleon-nucleon scattering and the usual interpretation is that 
pion production proceeds primarily by $\Delta$ production with 
the threshold for the reaction cross section being governed by 
the mass of the $\Delta$.

In the present reaction, the threshold behavior can perhaps be 
explained again by $\Delta$ production although the exchange of 
a single pion to form the $\Delta$ is not possible since a pion 
cannot couple to the kaon. Another possibility is for the 
reaction to proceed by the excitation of the {\it K}$^*$(892). The VPI 
group \cite{arndt1} used these two intermediate states to model 
the threshold behavior.  As can be seen from the insert in Fig. 
\ref{reaction} they obtain a rise in the reaction cross section 
very similar to that obtained by Martin, neither of which is 
sufficiently rapid. In the case of the VPI group \cite{arndt1} 
this is because the {\it K}$^*(892)$ has a higher threshold (1.077 
GeV/c in the laboratory) and, while the $\Delta$ has about the 
right threshold (0.886 GeV/c in the Lab), it has a significant 
width which causes the onset of the reaction cross section to be 
gradual. However, Oset and Vicente Vacas \cite{oset} had a 
reasonable success in reproducing the data by combining this 
mechanism with a chiral symmetry calculation.

Wyborny \ea \cite{wyborny} included final-state interactions 
between the $\Delta$ and kaon in the pion production channel and 
were able to produce a maximum in the {\it P}3/2 phase shift very 
similar to the one observed. Their result gives a shape closer 
to our results (or Martin's) than to those of Hashimoto 
\cite{hash} or Watts \ea \cite{watts} with which they compared. 
Figure \ref{wyborny} shows the comparison of their calculation 
with the four sets of phase shifts considered here.

In order to have a rapid onset for the reaction cross section 
through the mechanism of an intermediate particle production, that 
particle must have a small width. An interesting possibility is pion 
production by the intermediate step of $K+N\rightarrow \theta^++\pi$ 
where $\theta^+$ is the much discussed pentaquark. Hyodo \ea 
\cite{hyodo} have considered pion production in this reaction by 
this mechanism. The mass of the proposed pentaquark is thought to be 
around 1.54 GeV/c$^2$ and its width very small. The threshold for 
production by this means (0.758 GeV/c in the Lab) is slightly lower 
than the $\Delta$ threshold.  The diagrams for these three 
possibilities are given in Fig. \ref{diap1}.

The separation of the reaction cross section into its partial-wave 
components may be of some help in sorting out the reaction mechanisms (see 
Figs. \ref{reaction} and \ref{areact}). If one assumes that the 
intermediate particle is produced at rest in the center of mass then it 
should be in a relative s-wave. For the case of the $\Delta$ (with 
spin-parity 3/2+) in an s-wave, the only incident partial wave allowed is 
D3/2. It is interesting to note that in the present work, and in Martin's 
analysis, this partial wave has very little reaction cross section and the 
VPI analysis \cite{arndt1} has only a moderate contribution.  One possible 
explanation is that the production mechanism does not allow the formation 
of the $\Delta$-kaon final state in a relative s-wave. If we assume that 
it is produced in a relative p-wave then possibilities for the initial 
partial wave are {\it P}1/2, {\it P}3/2 and {\it F}5/2. We see that the 
{\it P}1/2 partial wave contributes a significant reaction cross section 
and the {\it P}3/2 partial wave also contains strength, although at 
slightly higher energies.

It is interesting to note that the {\it D}5/2 partial wave is important 
and, at least in the present work, dominates in the region 0.8 to 
1.0 GeV/c. The Martin analysis also gives the dominant waves to be 
{\it P}3/2, {\it P}1/2 and {\it D}5/2 (Fig. \ref{areact0}). In our work it 
seems to be the {\it D}5/2 partial wave which accounts for the rapid rise 
in the region 0.8 to 0.9 GeV/c. If one were to assume that this increase 
corresponds to the threshold for $\theta^+$ production in the s-wave 
then the $\theta^+$ would have to have spin-parity 5/2+.

Bland \ea \cite{blanda} measured the angular dependence of the pion 
production directly and found that $\Delta$ production is in the 
{\it P}1/2 and {\it P}3/2 waves. They attempted a partial-wave analysis of 
the production although it was only possible to include a limited number 
of amplitudes.  Their analysis indicated that the two {\it P} waves 
contributed about equally. They did not include the {\it D}5/2 wave in 
their study. In this same work \cite{blanda} a model of $\Delta$ 
formation by $\rho$ exchange was also presented which indicated that 
the expected ratio of {\it P}3/2 to {\it P}1/2 formation was 5:1, similar 
to what we find (Fig. \ref{reaction}).

Bland \ea \cite{blanda} also found strong evidence for interference 
with an even parity partial wave. This effect became much larger for 
low beam momenta and invariant masses away from the central mass of 
the $\Delta$. This even partial wave was only present in the final 
charge states K$^0$p$\pi^+$ and K$^+$n$\pi^+$ and not in the state 
{\it K}$^+$p$\pi^0$ where the kaon-nucleon system has isospin unity. Also 
observed in this paper was a strong asymmetry in the Daliz plot 
which, again, was not seen in the {\it K}$^+$p$\pi^0$ final state. These 
observations suggest that the asymmetry may be linked to the 
production of a {\it T}=0 {\it KN} final state, again consistent with the 
presence of an isospin 0 particle. They \cite{blanda} also studied 
the {\it K}$^*(892)$ production and were able to say that it occurred in a 
low angular momentum state consistent with our observation of a rise 
in the s-wave contribution in the region of the {\it K}$^*(892)$ 
threshold.

\subsection{Scattering Length}

The value of the {\it T}=1 scattering length has been very stable 
over the years. L\'evy-Leblond and Gourdin \cite{leblond} 
obtained a scattering length of --0.34 fm in what was probably 
the first analysis. Hyslop \ea \cite{hyslop} obtained a value 
of --0.33 fm, Cutkosky \ea \cite{cutkosky} found --0.28$ \pm$ 
0.06 fm and Martin \cite{martin} found --0.32 fm.

We obtain a central value of the s-wave scattering length of 
--0.308 fm from the best fit. One can estimate from the 
variation of minima in Table \ref{minima1} an error of $\pm$ 
0.002 fm. Fixing the scattering length at various values and 
re-fitting the rest of the parameters, one finds again an error 
of $\pm$ 0.002 fm. Adding the two errors in quadrature we can 
quote a value of --0.308 $\pm$ 0.003 fm. While this error is 
that obtained with the present data set, it is so small that one 
would have to expect that the addition of new data would lead to 
a change of the same order or larger. We note that the average 
from Refs. \cite{hyslop}, \cite{cutkosky} and \cite{martin} is 
--0.031 fm so our central value is what might be expected.

On the theory side, Barnes and Swanson \cite{barnes} obtained, 
in a quark Born approximation, a scattering length of --0.35 fm. 
When this value was corrected for unitarity, by solving for 
scattering from a potential, the value became --0.22 fm.

\subsection{Scattering Volumes}

The scattering volumes have always been estimated to be very small 
and are poorly determined. In early work, the p-waves were often 
neglected as input to the {\it T}=0 determinations. Cutkosky \ea 
\cite{cutkosky} obtained --0.04$ \pm$ 0.03 fm$^3$ for the {\it P}1/2 wave 
and 0.02 $\pm$ 0.02 fm$^3$ for the {\it P}3/2 wave, for example. The 
values obtained in the present work are very dependent on the 
minimum found, unlike the scattering length. For the higher values 
of $\chi^2$ the scattering volumes are small, in agreement with 
Cutkosky \ea \cite{cutkosky} but for the best fit, the volumes are 
somewhat larger in magnitude.

\begin{figure}[h] 
\epsfig{file=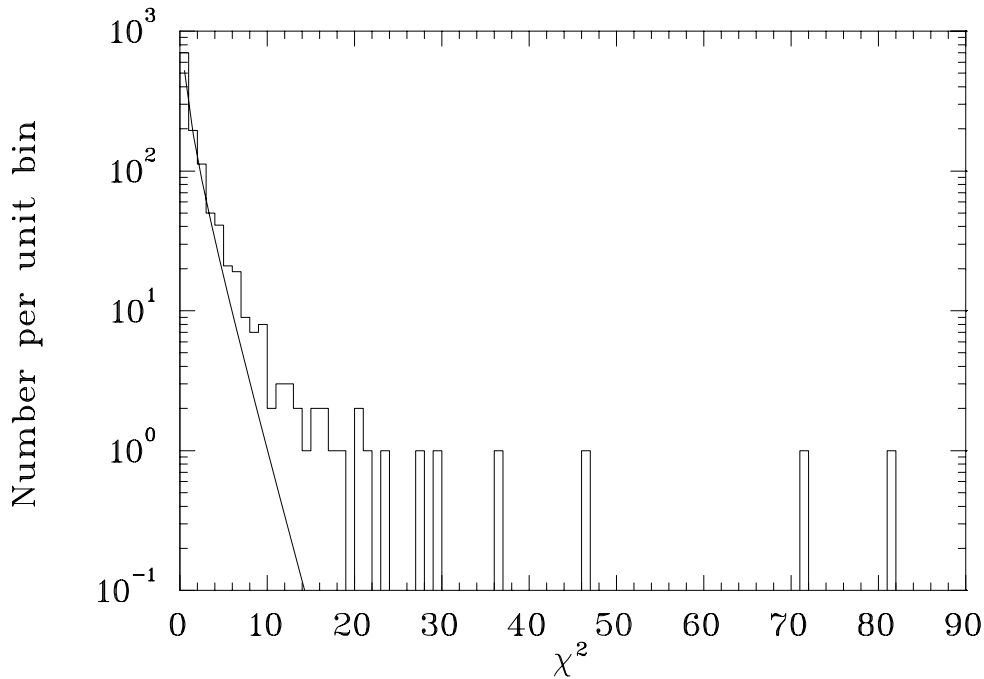,angle=90,height=3.5in} 
\caption{Distribution of the values of $\chi^2$ for 
isospin 0 for all data.} \label{chi2dis} \end{figure}

\begin{figure}[htb]
\epsfig{file=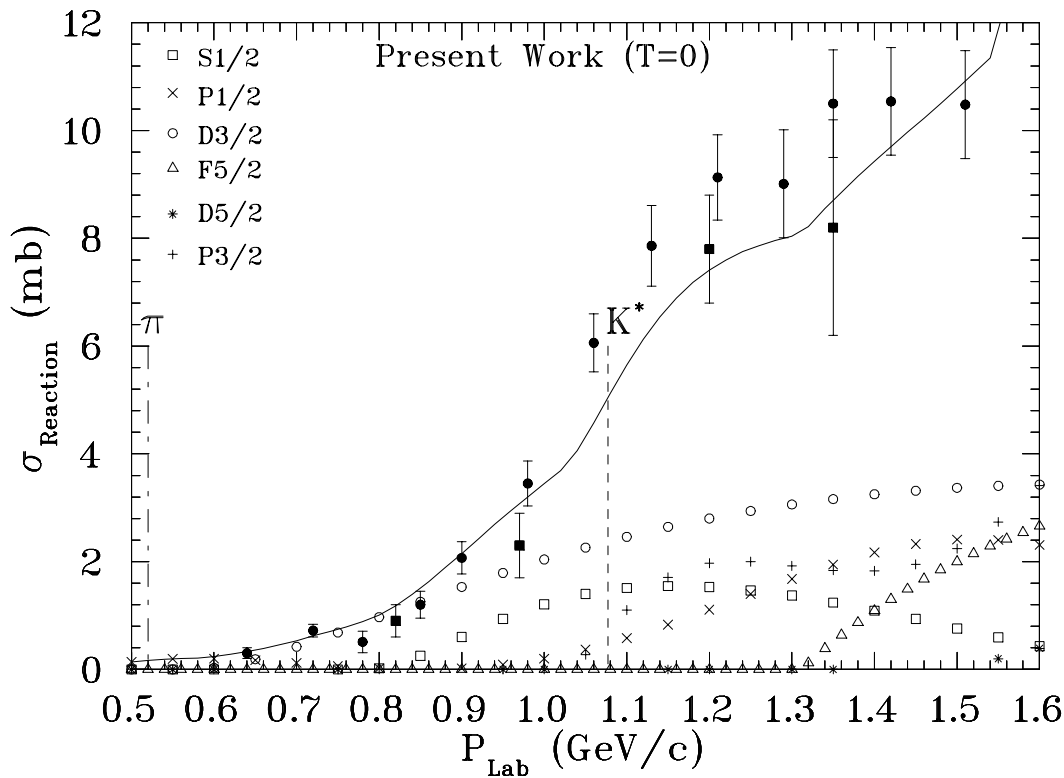,angle=90,height=4.in}
\caption{Partitioning of the {\it T}=0 reaction cross sections 
among partial waves for the present work with the standard 
reaction data.}
\label{gareact0} \end{figure}

\begin{figure}[htb]
\epsfig{file=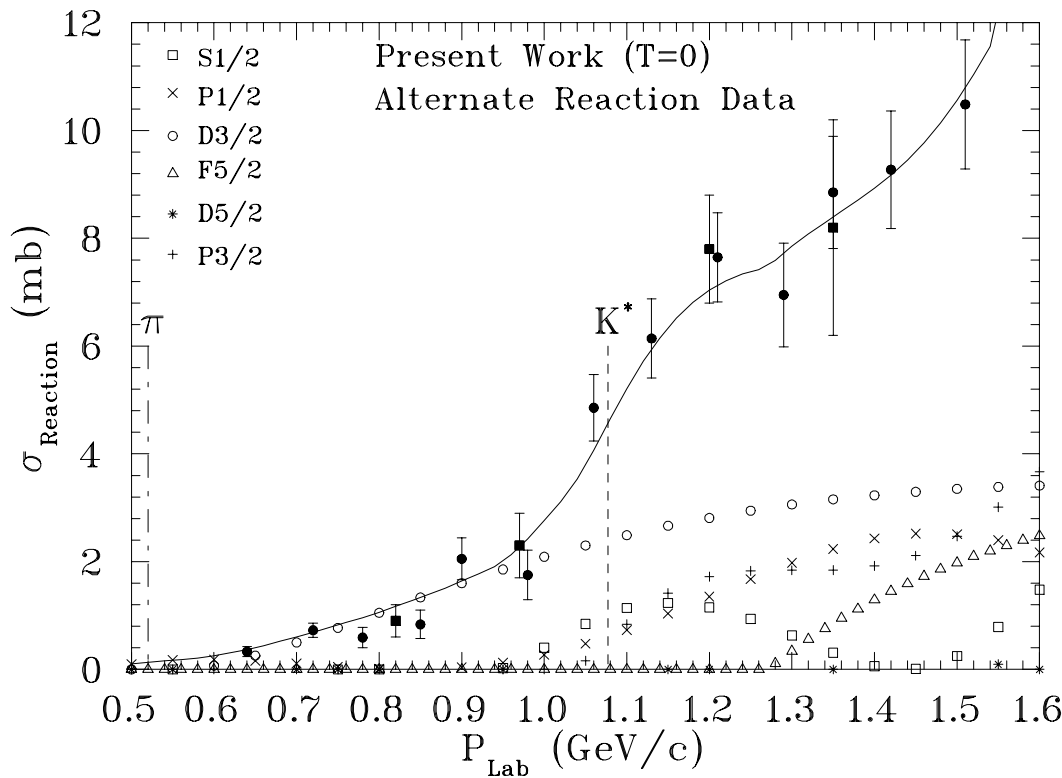,angle=90,height=4.in}
\caption{Partitioning of the {\it T}=0 reaction cross sections 
among partial waves for the present work with alternate reaction 
data.}
\label{gareact0ard} \end{figure}

\begin{figure}[htb] 
\epsfig{file=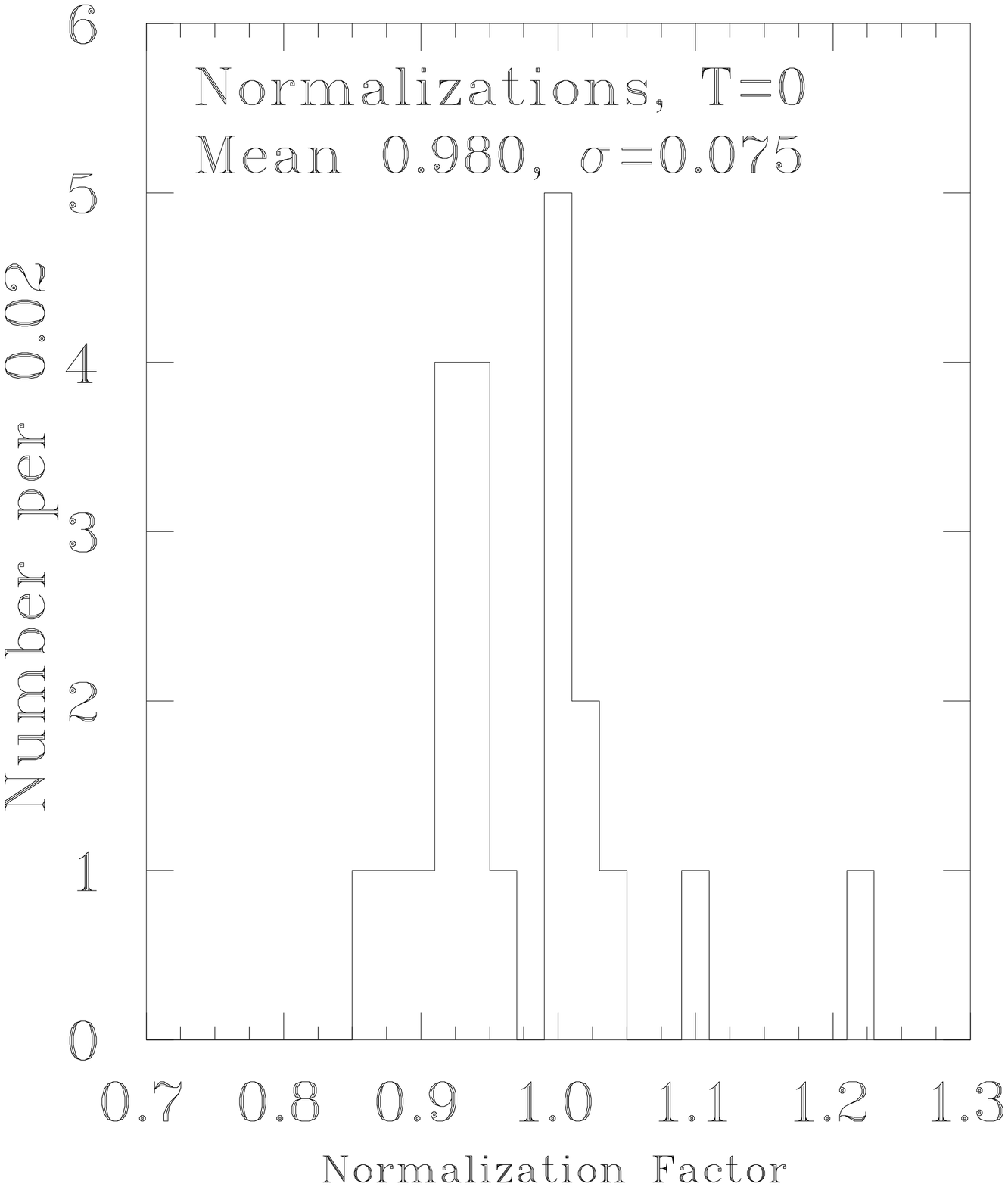,height=3.5in} 
\caption{Distribution of the normalization factors for 
isospin 0.} \label{norms0} \end{figure}

\begin{figure}[htb]
\epsfig{file=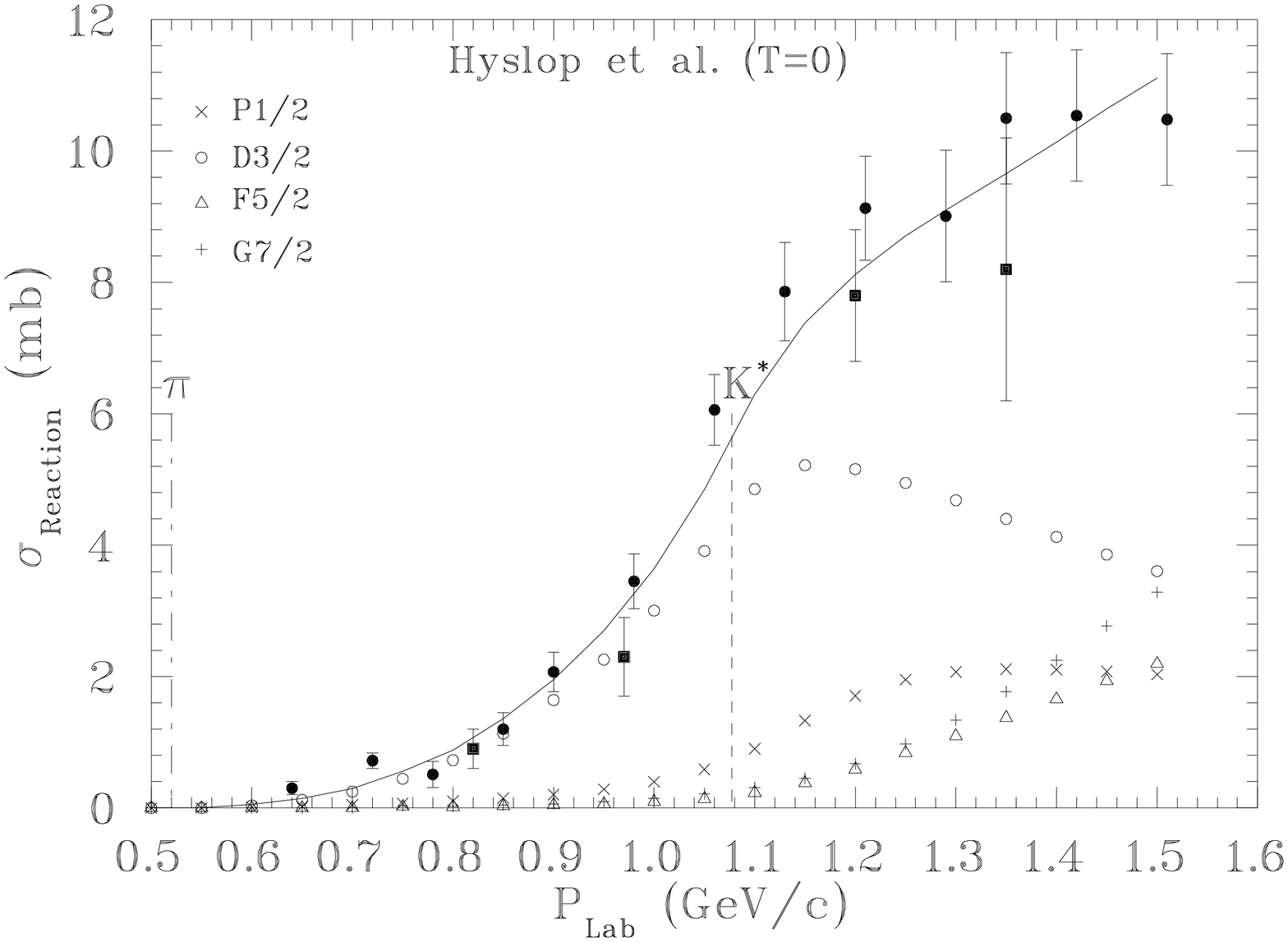,angle=90,height=4.1in}
\epsfig{file=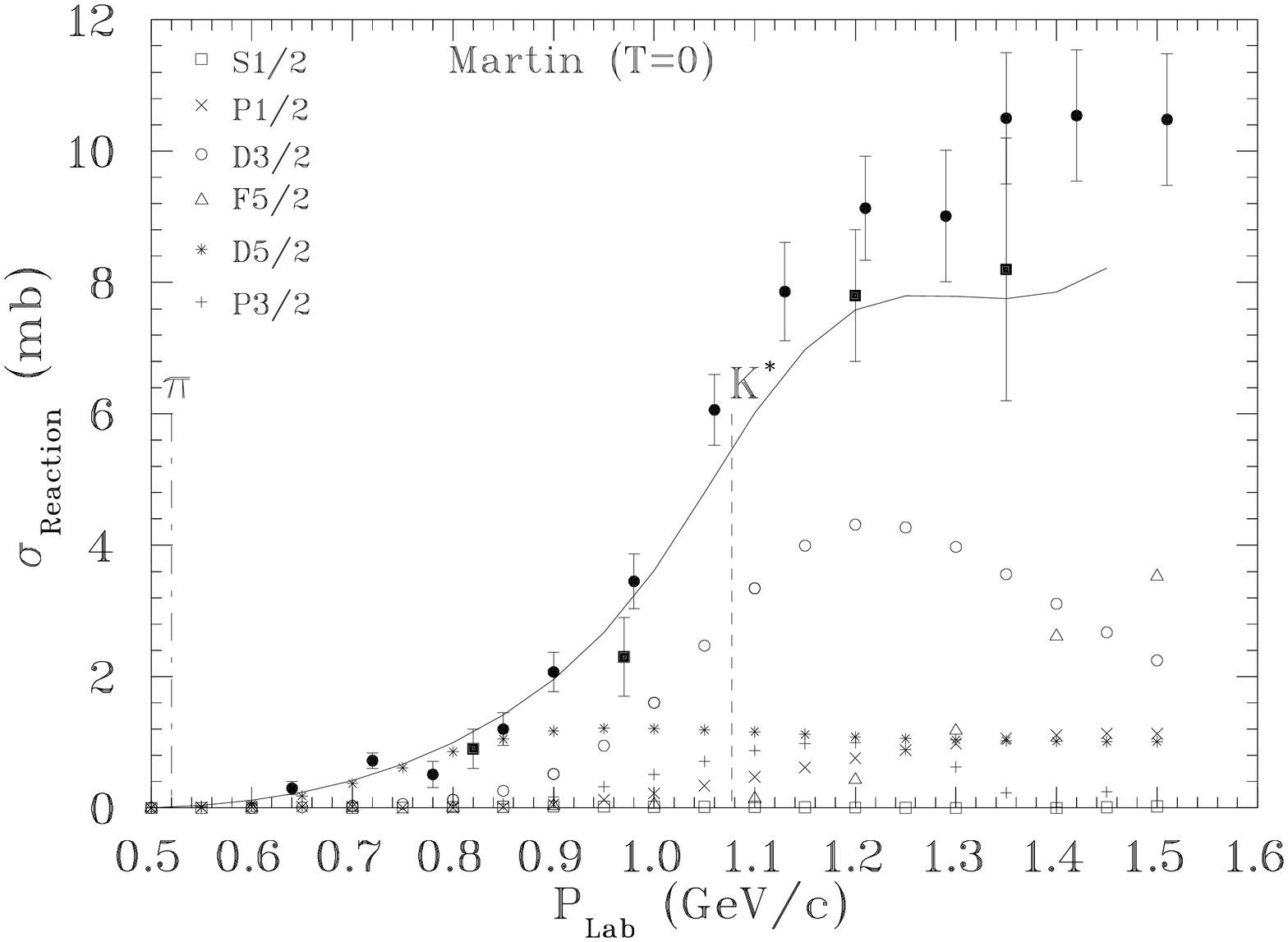,angle=90,height=4.1in}
\caption{Partitioning of the {\it T}=0 reaction cross sections among 
partial waves for Hyslop \ea \cite{hyslop} and Martin 
\cite{martin}.}
\label{areact0} \end{figure}

\begin{figure}[htb]
\hspace*{-.1in}\epsfig{file=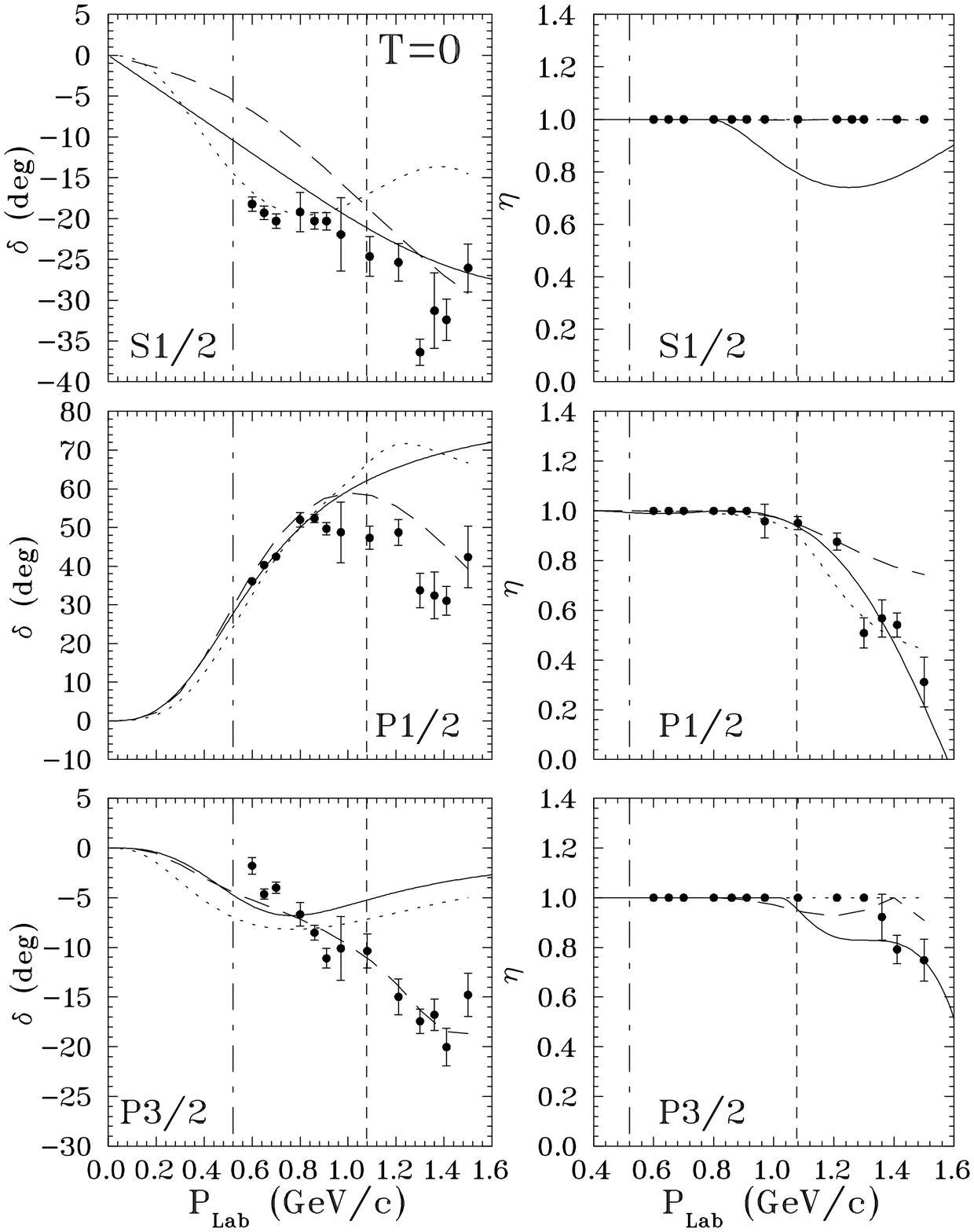,height=3.9in}
\epsfig{file=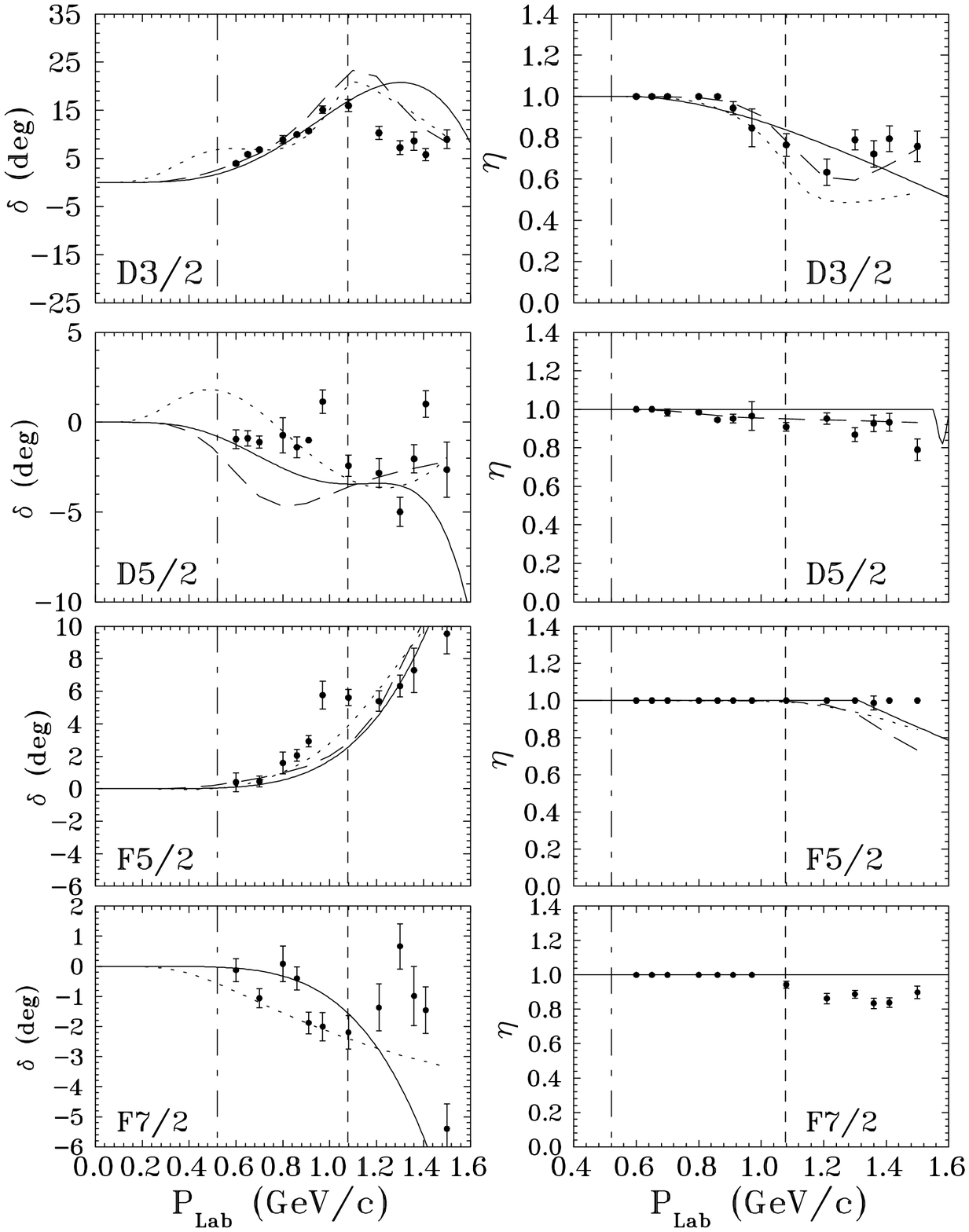,height=3.9in}
\caption{{\it T}=0 phase shifts and $\eta$'s obtained in the 
present work (solid line) compared with VPI 
\cite{hyslop} (dotted line), Hashimoto \cite{hash} 
(dots) and Martin \cite{martin} (dashed line). The 
vertical lines show the relevant thresholds, dash-dot: 
pion production threshold, dashed: threshold for production of  
a $K^+$.} \label{phases0} \end{figure}

\begin{figure}[htb]
\epsfig{file=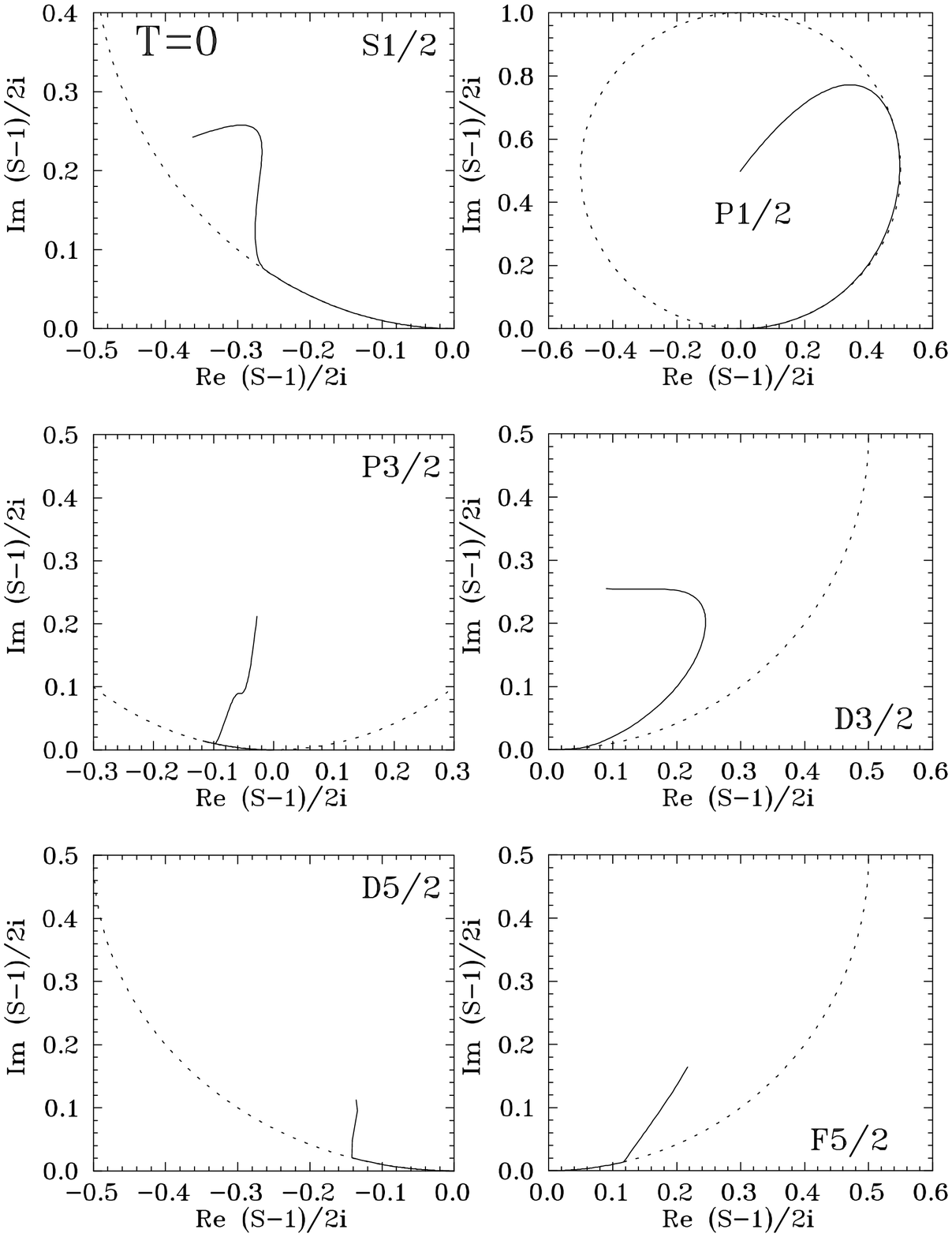,height=5.in}
\caption{Argand plots of the real and imaginary parts of
$\frac{S-1}{2i}$ for {\it T}=0.} 
\label{argand0} \end{figure}

\section{Analysis for {\it T}=0}

\subsection{Data Treatment}

The treatment of the isospin zero amplitude is more difficult 
since there exist no free neutron targets.  What is normally 
used is scattering from the neutron contained in the deuteron. 
Modern analyses \cite{kamalov,meissner} suggest that scattering 
from the meson cloud is not a problem. There are only slightly 
more than half of the number of data points compared to the {\it T}=1 
case and the reliability of the data is less, given that it must 
be extracted from deuteron data with corrections. Again most of 
the data can be obtained from compilations 
\cite{compilation,vpidb,pdg} but we try to give references to 
original papers \cite{cook,fisk,ray,giacomelli,hirata69, 
hirata71,adams,glasser,slater,carroll,sakitt,giacomellicex,giacomelliel, 
watts,robertson,damerell}. Normally the neutron result is 
obtained assuming that the interactions of the kaon with the 
proton and neutron are independent. This is usually a good 
approximation except at low energies. However, the {\it T}=1 reaction 
dominates the deuteron (for example the total cross section on 
the deuteron is $\frac{3}{2}\sigma (T=1)+\h \sigma(T=0)$) so a good 
knowledge of the {\it T}=1 amplitudes is needed. We 
used {\it T}=1 amplitudes from out best fit as input to the {\it T}=0 
determination.

The data used in our standard fit are all from incoherent {\it K}$^+$ 
deuteron scattering and $K^+$ deuteron total cross sections. There 
are 92 points of total cross section data, 17 points of one and two 
pion production data, 336 points of elastic cross sections from the 
neutron, 43 points of elastic polarization from the neutron, 657 
points of charge exchange cross section data and 44 points of charge 
exchange polarization data for a total of 1189 points.

\begin{table}
$$ \begin{array}{|c|c|c|c|c|c|c|}
\hline
{\rm Case}&\chi^2&N&\chi^2/N&{\rm Scatt.\ Len.\ (fm)}
&{\rm P1/2\ (fm}^3{\rm )}&{\rm P3/2\ (fm}^3{\rm )}\\
 \hline
{\rm Best}  &  1670.561 &1170&1.428 &-0.1048&0.183&-0.029 \\
 \hline
a   & 1671.214  &1170&1.428 &-0.1015&0.183&-0.028 \\ 
 \hline
b   & 1670.561 &1170&1.428 & -0.1053&0.182&-0.029 \\  
 \hline
c   & 1670.758&1170&1.428 & -0.1055&0.182&-0.029\\  
 \hline
d   & 1671.085 &1170&1.428 &-0.1055&0.183&-0.030\\   
 \hline
e  &  1675.274&1170&1.432 & -0.1067&0.181&-0.022 \\ 
 \hline
f  &  1675.301  &1170&1.432 & -0.1003&0.181&-0.022\\  
 \hline
 \hline
\chi^2<10&1540.480&1162&1.326&-0.1027&0.182&-0.028\\ 
 \hline
{\rm No\ Krauss/Weiss}& 1649.073&1155&1.428&-0.0988&0.177&-0.027\\
 \hline
{\rm No\ Double} & 1662.132&1170&1.421& -0.1166&0.200&-0.038\\ 
 \hline
{\rm Alter.\ Reaction}& 1665.105&1170&1.423&-0.0957&0.183&-0.021\\
 \hline
{\rm Stenger\ Full} & 1714.224 &1190&1.441&-0.1096&0.174&-0.036 \\
 \hline
{\rm Stenger\ Partial} &1692.632&1186&1.427 &-0.1036&0.181&-0.029 \\
 \hline
{\rm No\ Damerell}&1284.174 &1008&1.274&-0.0997&0.205&-0.025\\ 
 \hline
{\rm K0p}&1895.322&1319&1.437&-0.1069&0.173&-0.032\\ 
 \hline
\end{array}
$$
\caption{Values of $\chi^2$ for the best fit and the variations 
made in the present work for {\it T}=0. Cases a-f show different fits 
obtained with the full code. Case ``No Krauss/Weiss'' 
corresponds to a fit in which the data of Refs. \cite{krauss} 
and \cite{weiss} were left out. In case ``No Double'' the 
double scattering correction was omitted and in case ``Alter. 
Reaction'' the reaction data was replaced by the average as 
explained in the text. The case ``Stenger Full'' gives the 
results with the full Stenger data \cite{stenger} included and 
case ``Stenger Partial'' the results where the most forward 
points were omitted. For the case ``No Damerell'' the Damerell 
\ea data \cite{damerell} were omitted from the fit. For the 
line labeled ``K0p'' the data of Armitage \ea \cite{armitage} 
were included in the fit.
\label{minima0}}
\end{table}

\begin{table}
$$ \begin{array}{|c|c|r|r|c|c|}
\hline
LJ&q_R\ (GeV/c)&\gamma_1\ \ \ \ &\gamma_2\ \ \ \ &\gamma_3&{\rm 
Equation}\\
 \hline
 S1/2 &  0.4350   &3.581   &-4.349 &&\ref{etad32}\\
 \hline
 P1/2 & 0.2515 &-0.689  &0.822 &&\ref{etad32}\\
 \hline
 P3/2 & 0.5303 &-9.101  &46.593  
&-79.735&\ref{etad32} \\
 \hline
 D3/2 &0.3351  &0.887 &&&\ref{etad32}\\
 \hline
 D5/2 &0.7120 & & &&\ref{etad32}\\
 \hline
 F5/2  & 0.6352  & 1.485 &&&\ref{etas}\\
 \hline
\end{array}
$$
\caption{Parameters for the representation of the inelasticity,
$\eta_{\ell,j}$ for {\it T}=0 using the form of Eq. \ref{etad32} for all
the partial waves except the {\it F}5/2 wave which uses Eq. \ref{etas}.
The threshold for the {\it D}5/2 wave is P$_{Lab}$=1.55 GeV/c
so the $\eta$ for this partial wave can be taken as unity over our fitted 
range.
\label{pars0eta}.}
\end{table}

\begin{table}
$$ \begin{array}{|r|r|r|r|r|c|}
\hline
LJ&a\ {\rm (GeV/c)}^{-(2\ell+1)}&b_{1,\lpm}\ {\rm 
(GeV/c)}^{-2}&b_{2,\lpm}\ 
{\rm (GeV/c)}^{-4}&b_{3,\lpm}\ {\rm (GeV/c)}^{-6}&{\rm Equation}\\
 \hline
 S1/2 &   -0.531  &-1.206  &1.362  &&\ref{spwaves}\\
 \hline
 P1/2 & 23.765  &3.690 &&&\ref{spwaves}\\
 \hline
 P3/2 &   -3.808  &2.919   &-10.042  
&212.021&\ref{spwaves} \\
\hline \hline
LJ&c_{\pm}\ {\rm (GeV/c)}^{-5}&d_{\pm}\ {\rm (GeV/c)}^{-7}&e_{\pm}\ 
{\rm (GeV/c)}^{-9}&&{\rm Equation}\\
\hline
 D3/2 & 12.548 & -22.412 &&&\ref{dwaves}\\
 \hline
 D5/2 & -7.528 &32.262  &-37.422 &&\ref{dwaves}\\
\hline \hline
LJ&f_{\pm}\ {\rm (GeV/c)}^{-7}&&&&{\rm Equation}\\
 \hline
 F5/2  & 2.836&&&&\ref{fwaves}\\
 \hline
 F7/2 & -1.731&&&&\ref{fwaves}\\
\hline
\end{array}
$$
\caption{Parameters for the representation of the 
phase shifts for {\it T}=0 using the form of Eq. \ref{spwaves}
for the {\it S} and {\it P} waves, Eq. \ref{dwaves} for the {\it D}
waves and Eq. \ref{fwaves} for the {\it F} waves.}\label{pars0del}
\end{table}

As might be expected, the pruning of the {\it T}=0 data is more 
significant and problematic than the {\it T}=1 case. In order to 
choose which points to eliminate, a preliminary fit was made 
with all data included. By binning the values of $\chi^2$ (bins 
of unit size were chosen) a distribution was obtained. This 
distribution, along with the expected distribution obtained from 
a $\chi^2$ distribution with one degree of freedom is shown in 
Fig. \ref{chi2dis}. We see that the expected number of counts in 
a bin of unit size falls to 0.1 around $\chi^2$=15. For this 
reason, the 19 points with $\chi^2\ge 15$ were removed leaving 
1170 data points to be fit. It is clear that there is still an 
excess of points below $\chi^2=15$ but it is impossible to tell 
which points to remove so no further pruning was done and we 
must expect to have a higher $\chi^2$. However, we did make one 
run with all points giving a $\chi^2>10$ being removed. The 
results of this fit are shown in Table \ref{minima0}.

 \subsection{Double Scattering}

At very low energies, double scattering in the deuteron gives an 
important contribution to the total cross section. Other 
observables are not sufficiently well measured that it will have 
a significant effect.

The double scattering amplitude at 0$^\circ$ is given by \cite{book}
\eq
f_D(\theta=0)=f_D(\bfk,\bfk)=\frac{1}{2\pi^2}\int \frac{d\bfq
f_b(\bfq,\bfk)f_a(\bfk,\bfq)}{q^2-k^2-i\epsilon}
z\left(|\bfk-\bfq |\right)
\qe
where $\bfk$ is the initial and final (on-shell)
momentum of the scattering meson, $z(\bfp)$ is the two-body
form factor, and $f(\bfk,\bfq)$ and $f(\bfq,\bfk)$ are
half-off-shell basic scattering amplitudes.   
   
For the scattering amplitudes we write the off-shell 
dependence as
\eq
f(\bfq,\bfq')=f_0v(q)v(q');\ \ v(k)=1 \ \ 
{\rm where\ the\ form\ \ \ }
v(q)=\left(\frac{k^2+\Lambda^2}{q^2+\Lambda^2}\right)^2
\qe
will be assumed. 

\eq
f_D(\theta=0)=
\frac{ik}{4\pi}\int d\Omega_{\bfq}
f_b(\bfq,\bfk)f_a(\bfk,\bfq)
z\left(|\bfk-\bfq|\right)+
\frac{1}{2\pi^2}{\cal P}\int \frac{d\omega_{\bfq}q^2 dq
f_b(\bfq,\bfk)f_a(\bfk,\bfq)}{q^2-k^2}
z\left(|\bfk-\bfq|\right)
\qe

We will consider only the s- and p-waves for this correction. The 
isospin 1 s-wave is the strongest so we consider the double 
scattering between it and the s- and p-waves of the neutron.
With these assumptions
 
$$ f_D(\theta=0)=
\frac{ikf_0^pf_0^n}{4\pi}\int d\Omega_{\bfq}
z\left(|\bfk-\bfq|\right)+
\frac{f_0^pf_0^n}{2\pi^2}{\cal P}\int \frac{d\omega_{\bfq}q^2 dq
}{q^2-k^2}z\left(|\bfk-\bfq|\right)
$$
\eq
+\frac{ikf_0^p f_1^n}{4\pi}\int d\Omega_{\bfq} x
z\left(|\bfk-\bfq|\right)+
\frac{f_0^p f_1^n}{2\pi^2}{\cal P}\int \frac{d\omega_{\bfq}q^2 
dq x}{q^2-k^2}z\left(|\bfk-\bfq|\right)
\qe
where $x$ is the cosine of the angle between $\bfq$ and $\bfk$. 
The double scattering contribution to the total cross section
will be
$$
\sigma_T= 2{\rm Im} \left[
if_0^pf_0^n\int d\Omega_{\bfq}z\left(|\bfk-\bfq|\right)+
\frac{2f_0^pf_0^n}{\pi k}{\cal P}\int \frac{d\omega_{\bfq}q^2 dq
}{q^2-k^2}z\left(|\bfk-\bfq|\right)\right.
$$
\eq \left.
+if_0^p f_1^n\int d\Omega_{\bfq} xz\left(|\bfk-\bfq|\right)+
\frac{2f_0^p f_1^n}{\pi k}{\cal P}\int \frac{d\omega_{\bfq}q^2 
dq x
}{q^2-k^2}z\left(|\bfk-\bfq|\right)\right]
\qe
 
Where the factor of 2 comes from the fact that there are two
orders of scattering possible.
To include charge-exchange scattering we can replace
\eq
f_0^pf_0^n\longrightarrow f_0^pf_0^n-\h f_{0x}^2
\qe

Near zero energy the amplitudes become real (so the principal value 
terms become very small) and the p-wave amplitude goes to zero so 
that only the first term remains and gives a contribution of $8\pi 
f_0^p f_0^n$. Since the fourth term is small for both reasons we 
have neglected it. The function $z$ was computed using the 
one-pion-exchange deuteron wave function \cite{oped} which gives a 
good representation of the momentum distribution in the deuteron 
\cite{dg}. While we believe that this correction is needed to get a 
proper fit, the results with it being left out are shown in Table 
\ref{minima0}.

\subsection{Reaction Cross Sections}

Aside from the values given by Hirata \ea \cite{hirata71}, the 
principal reaction cross sections (one and two pion production) 
are from Giacomelli \ea \cite{giacomelli}. They used their 
deuteron and proton pion production data to extract the neutron 
data. A more consistent way for us is perhaps to use our fit to 
the proton results with their deuteron production data to 
obtain directly the {\it T=0} reaction cross section. These can be 
obtained directly in the single scattering impulse 
approximation from the following equations.

\eq 
\sigma_R({\rm T=1})=\sigma_R({\rm 
proton});\ \ \ \sigma_R({\rm deuteron})= \sigma_R({\rm 
proton})+\sigma_R({\rm neutron}) =\frac{3}{2}\sigma_R({\rm T=1})+\h 
\sigma_R({\rm T=0}) 
\qe 

With not much to choose between the two one might hope to 
improve the errors by averaging the two determinations. Figures
\ref{gareact0} and \ref{gareact0ard} show the result of a fit with 
the original reaction data and the alternate reaction data.
Table \ref{minima0} shows the effect of using this 
alternative reaction data.

\subsection{Normalizations}

We maintained the constant error of 3\% for most of the 
normalizations although, for the reasons mentioned above, we 
might expect it to be larger. Figure \ref{norms0} shows the 
distribution of normalizations. The distribution is wider than 
for the {\it T}=1 case and there is a slight bias. The calculated 
width of the distribution is influenced by one outlier from the 
Damerell \ea elastic data \cite{damerell}.

\begin{figure}[htb]
\hspace*{-.1in}\epsfig{file=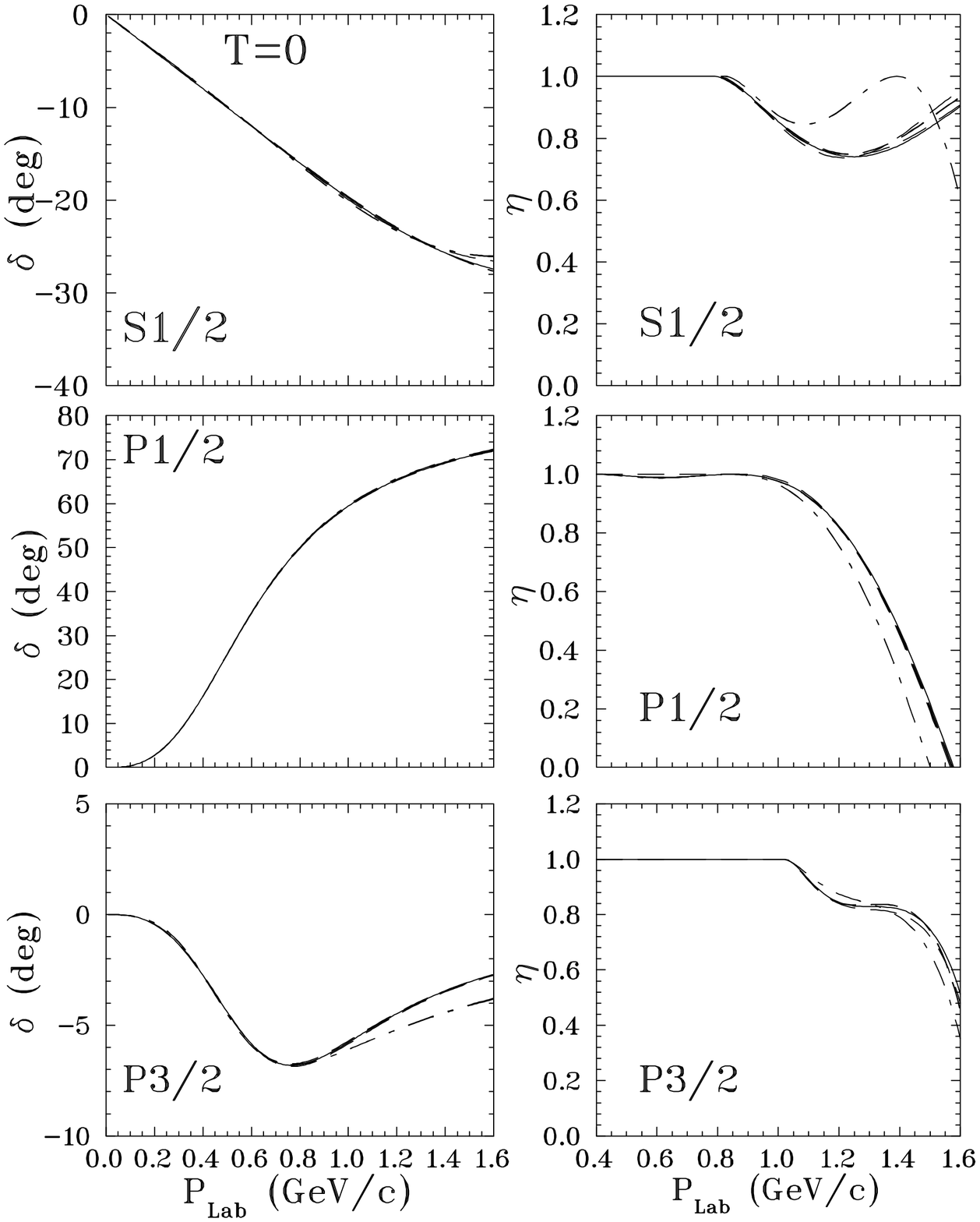,height=3.9in}
\epsfig{file=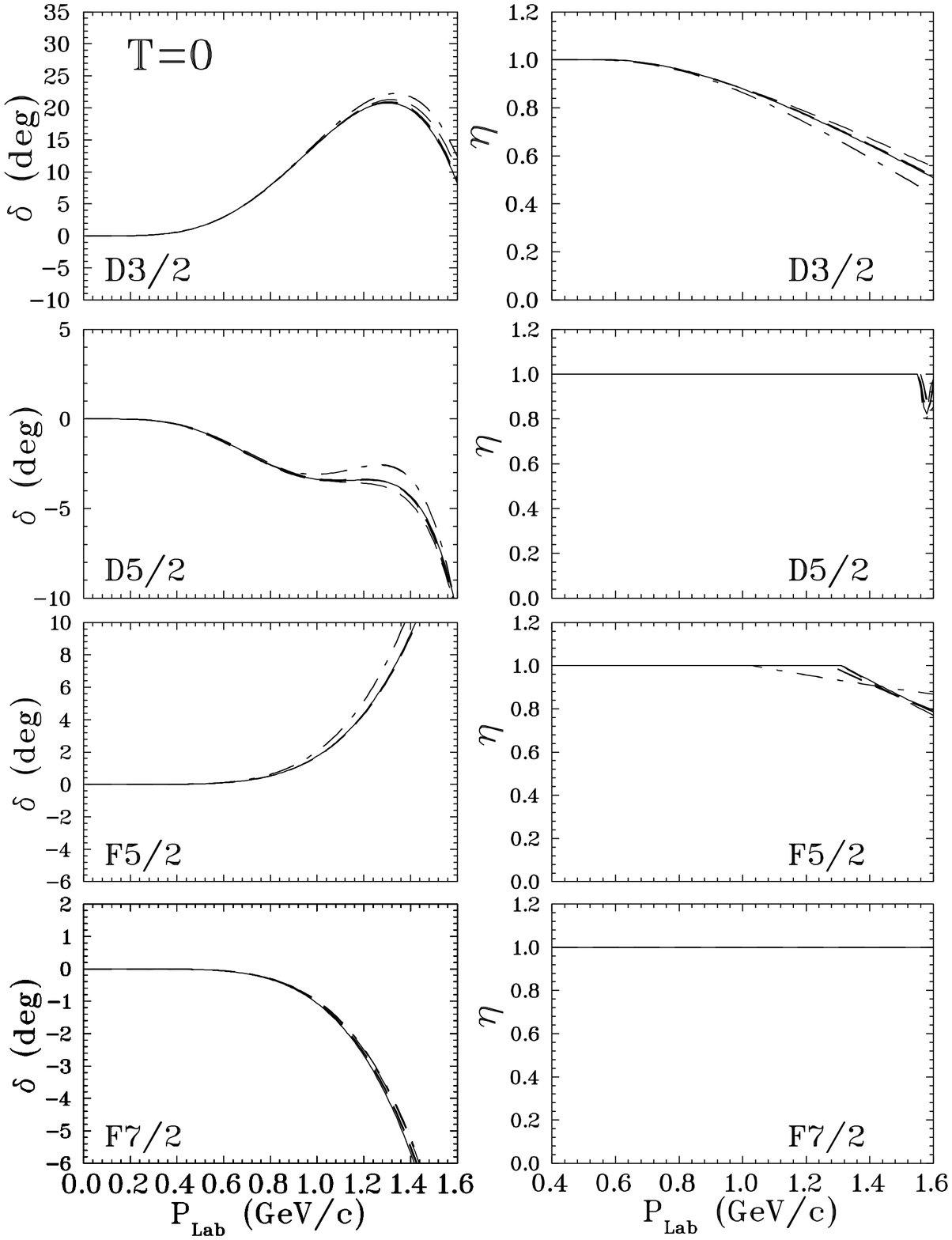,height=3.9in}
\caption{T=0 phase shifts and $\eta$'s obtained in the present 
work (solid line) compared with the variations. The solid line 
is the best fit, the dashed lines correspond to fits a, b, 
c, and d, and the dash-dot lines to the poorer fits e and f where the
labeling of the cases are indicated in Table \ref{minima0}.} 
\label{compare0} \end{figure}

\begin{figure}[htb]
\epsfig{file=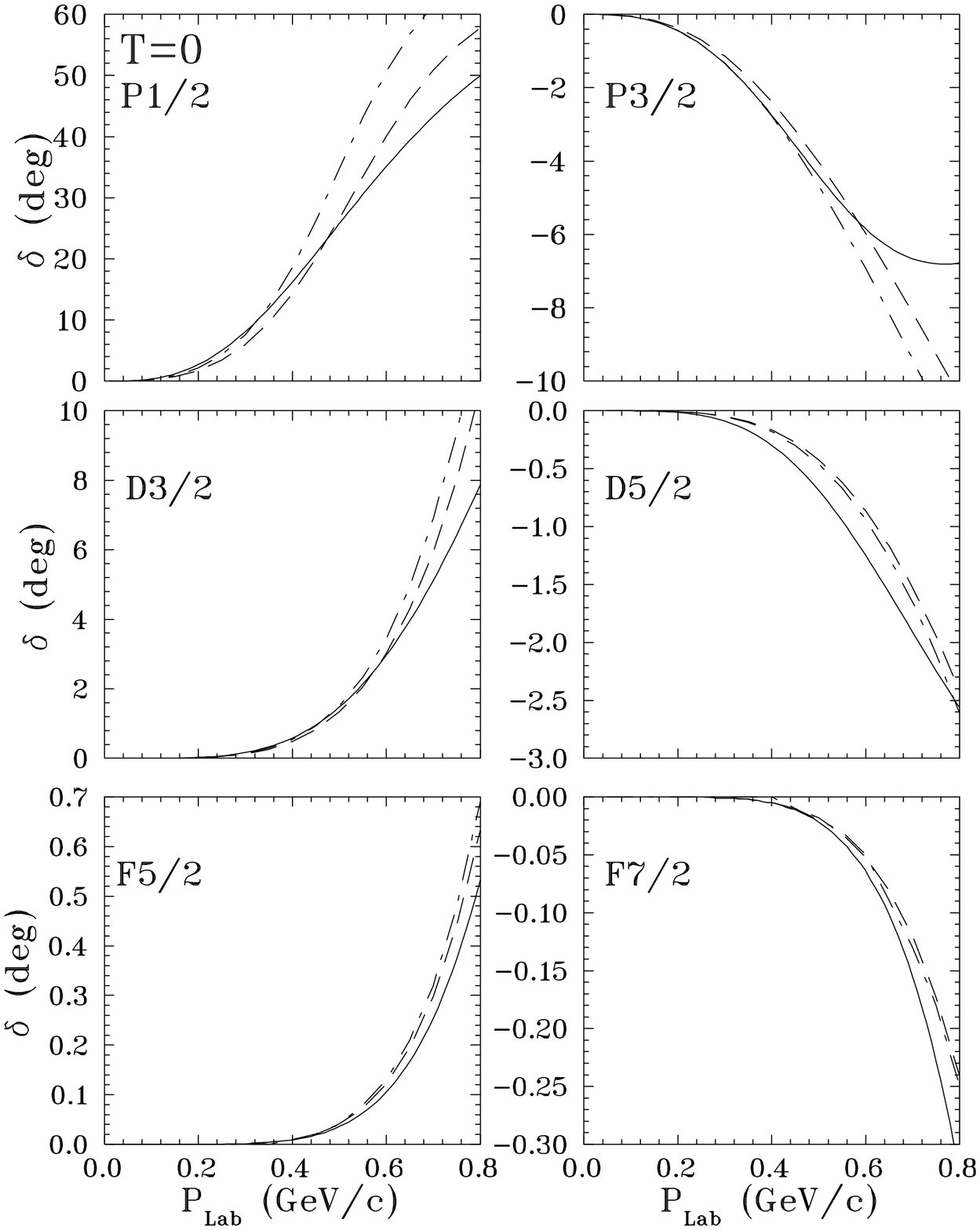,height=5.in}
\caption{Square-well spin-orbit model. The solid line gives
the results of the present amplitude analysis, the dotted lines 
the model described in the text without a central term and the
dash-dot lines the model with the central potential which gives
the correct s-wave scattering length.} 
\label{square} \end{figure}

\subsection{Other Work}

\subsubsection{Analysis of Hashimoto}

Hashimoto \cite{hash} observed a resonance-like structure in 
the {\it T}=0 {\it D}3/2 wave which had been seen before 
\cite{martin,nakajima,kato}. While we see a loop in the Argand 
diagram, it is less pronounced than what he saw.

\subsubsection{Analysis of Martin}

Martin's database was somewhat smaller than the one used here. 
He used, in addition to the direct deuteron data, real parts of 
the amplitude obtained from a dispersion relation analysis 
using the K$^+$ and K$^-$ total cross sections. The dispersion 
relations were once subtracted with the subtraction point was 
taken at zero energy which means that the scattering lengths 
were input. He assumed that the isospin 0 scattering length was 
$0\pm 0.04 $ fm based on previous analyses. His fit resulted 
in a scattering length of --0.035 fm in agreement with his 
small input value.

We inserted Martin's solution in our code to compare with our 
database. Since he had a smaller amount of data and the 
parameters were not fit to the present data base, one cannot 
expect a reduced $\chi^2$ very close to what he obtained. We 
find (for 1170 points) a $\chi^2$ of 2849 or a ratio of 2.44. 
for the full data set. Restricting the comparison to data below 
1.2 GeV/c (because of the missing high partial-waves problem 
mentioned before) we find a $\chi^2$ of 1430 for 827 points for 
a ratio of 1.73. His original fit (number 2) found a $\chi^2$ of 
924.5 for 760 points or 1.22.

\subsubsection{Analysis of the VPI Group}

Hyslop \ea \cite{hyslop} use $\Delta$ production as a model for 
inelasticity in spite of the fact that the K$^+$$\Delta$ final 
state is forbidden in the {\it T}=0 channel. Comparing their amplitudes 
with our data base we find a $\chi^2$ of 2776 on 1170 points for a 
ratio of 2.37. They found 3181 for 1746 data points for a ratio of 
1.82. Again, we remark that their parameters were not adjusted to 
our data base and phase shifts of only limited accuracy are 
available so the same value cannot be expected.

\section{Results for {\it T}=0}

Figure \ref{phases0} shows the phase shifts obtained for the {\it T}=0 
fit compared with three other analyses and Tables \ref{pars0eta} 
and \ref{pars0del} show the parameters which give the phase 
shifts and inelasticities. Figure \ref{argand0} shows the 
behavior of the {\it T}=0 phase shifts in Argand plots. There is some 
structure in the plots but nothing that can surely be associated 
with resonances.

\subsection{Reaction Channels}

The isospin 0 channel is quite interesting from the point of 
view of the mechanism for pion production. Because of isospin, 
$\Delta$ production is not allowed, nor is the production of the 
isoscalar $\theta^+$. Indeed, the reaction cross section is seen 
to be smooth in the region of these two thresholds, unlike the 
{\it T}=1 case. The {\it K}*(892) production is permitted and one does 
observe a rapid rise around where it would be expected (Figs. 
\ref{gareact0} and \ref{gareact0ard}). This is particularly true 
of our results but is consistent with the reaction cross 
sections of the VPI group and Martin as is seen in Fig. 
\ref{areact0}. Since the {\it K}$^*$(892) has spin-parity 1$^-$, if it 
is produced with the nucleon in a relative s-state, the 
spin-parity values possible are 1/2$^-$ and 3/2$^-$ which 
correspond to the incident waves of {\it S}1/2 and {\it D}3/2. Indeed, we 
see the {\it S}1/2 wave giving an important contribution to the 
reaction cross section at the threshold and the {\it D}3/2 wave is the 
largest single contributor. Somewhat surprising is the dominance 
(or at least importance) of the {\it D}3/2 channel. This is the 
partial wave expected for the (isospin forbidden) production of 
a $\Delta$ in the s-wave. However, this wave may be simply the 
dominant wave for the non-isobar production as well as receiving 
a contribution for the {\it K}$^*(892)$ production.

\subsection{Scattering Length}

The {\it T}=0 scattering length has a rather checkered history.  
L\'evy-Leblond and Gourdin \cite{leblond} obtained a value of --0.05 
fm, with large but unspecified errors. Stenger \ea \cite{stenger} 
found a value of +0.04 fm (a value which was commonly used in 
dispersion relation work \cite{martinrome}). There were also 
suggestions that it might be positive and large \cite{alcock}. 
Presumably based on this previous work, Martin \cite{martin} set the 
scattering length to zero with an error of 0.04 fm as the 
subtraction point for his dispersion relation constraint. Martin's 
fit gives --0.035 fm, although he states that it has a large error.  
Later Martin gives \cite{martinpl} a value of 0.02 fm and then in 
still later work \cite{martin78} he found --0.23$\pm$0.18 fm.

Barnes and Swanson \cite{barnes} obtained a theoretical 
scattering length of --0.12 fm from a Born quark model. To 
compare with the experimental value they performed their own 
extrapolation to zero energy based on single-energy analyses 
and found --0.09 or --0.17 fm depending on which analysis they 
used.

From Hyslop \ea we extrapolate a value of --0.019 although they 
quote in the paper a value of zero. It can be seen from Fig. 
\ref{phases0} that the behavior of the isospin 0 s-wave in their 
fit is rather different from ours and has a great deal of 
variation in the low-energy region where there are no data. It 
can only be assumed that the variation in this case is a result 
of a fit at higher energies. It is seen that the trend of the 
curve above 1.1 GeV/c is noticeably different from the other 
determinations.

For the {\it T}=0 s-wave scattering length we adopt for the central value 
our best fit value of --0.105 fm from Table \ref{minima0}. The 
error for the uncertainty in the minimum is estimated from Table 
\ref{minima0} to be 0.002 fm. In order to estimate the statistical 
error the fit was redone for several fixed values of the scattering 
length varying all other parameters. From the resulting $\chi^2$ 
curve the error can be estimated to be 0.01 fm. For values of 
scattering length close to the central value, the $\chi^2$ curve is 
symmetric but for larger deviations it is not, rising steeply for 
small values. For the value of --0.035 fm given by Martin 
\cite{martin}, the $\chi^2$ corresponds to 8.6 standard deviations 
from our central value. Since the statistical error dominates, we 
take the scattering length as $-0.105\pm 0.01$ fm. This error does 
not include possible errors from the variations in the database. 
Observations on the change in value from the omission of data sets 
as shown in Table \ref{minima0} suggest that the error from this 
source can be expected to be of the same order or slightly 
smaller.

\subsection{Scattering Volumes}

The {\it P}1/2 scattering volume is well determined by the fit to be 
0.183 $\pm$ 0.005 fm where the error comes from an examination of 
Table \ref{minima0} assuming that there is no reason to exclude the 
double scattering correction or the Damerell data.

The {\it P}3/2 scattering volume is smaller and more poorly 
determined. Again from Table \ref{minima0} we take the value of 
--0.029 $\pm$ 0.008 fm.

\subsection{Spin-orbit Splitting}

The phase shifts for $\ell>0$ shown in Fig. \ref{phases0} 
display a remarkable symmetry below threshold. All phase shifts 
for $j=\ell-\h$ are positive and all of those for $j=\ell+\h$ 
are negative. In order to see how far a pure spin-orbit 
interaction would go toward explaining this behavior, we 
calculated a simple model consisting of scattering from a 
square well potential with strength $V$ where
\eq
V=V_0 {\bf L}\cdot {\bf S}=\left\{\begin{array}{cc}
\h V_0\ell& j=\ell+\h\\
-\h V_0(\ell+1)&j=\ell-\h
\end{array}\right.
\qe

The radius of the well was taken to be $R=0.85$ fm (corresponding 
to an rms radius of 0.66 fm). and the strength, $V_0$ was chosen 
to be 0.36 GeV. The results are shown in Fig. \ref{square} with the 
dashed line. The angular momentum barrier changes a great deal from 
one value of $\ell$ to another and the potential strength also 
changes over a significant range. The rather remarkable agreement 
indicates that for $\ell>0$ the phase shifts in the lower energy 
region are described by a pure spin-orbit interaction.

Such a potential gives no contribution to the s-wave. If we 
introduce a central potential (independent of $\ell$) in all 
partial waves of strength 0.04 GeV we obtain an s-wave scattering 
length of --0.11 fm (in agreement with our determination from the 
data). The result for the higher partial waves is shown in Fig. 
\ref{square} by the dash-dot lines. Thus, including a central 
potential of sufficient strength to give the moderate s-wave
scattering length does not destroy the good agreement seen before.

\subsection{Variations in the fit}

As in the {\it T}=1 case, a number of different minima were found 
corresponding to different starting points. The basic properties 
of the different fits are given in Table \ref{minima0} (caes 
a-f) and the variations of the phase shifts are shown in Fig. 
\ref{compare0}.

The elastic data by Stenger \ea \cite{stenger} was not included 
in the general fit. These data were among the first to find a 
very small scattering length for the {\it T}=0 channel. They consist 
of charge exchange and elastic scattering from the deuteron. In 
the elastic scattering there was no separation of scattering 
from the neutron or proton or, indeed, coherently from the 
deuteron (leaving it intact). Thus, for the elastic scattering 
the cross section from the proton should be added to that of 
the neutron and coherence must be taken into account as well. 
The contamination from coherent scattering is largest for small 
angles.  For the charge-exchange cross section small angle 
scattering (small momentum transfer) tends (without spin-flip)
to leave two protons in a triplet s-wave state. Since this state 
is blocked by the Pauli principle and the spin-flip amplitude 
is small at small angles, this effect leads to a very large 
suppression of the charge-exchange cross section such that it 
is far from charge exchange on a free neutron. The data were 
taken in large angle bins (0.4 in $\cos\theta$). In order to 
estimate the effect of leaving out this data set we made two 
runs, one in which the two lowest energies were fully included 
in the fit and one in which the most forward points (at 
$\cos\theta=0.8$) were excluded. The results are summarized in 
Table \ref{minima0}.

There are two modern data sets of total cross section data, 
those of Krauss \ea \cite{krauss} and Weiss \ea \cite{weiss}. 
In the fitting process these data suffer a renormalization 
(down) of about 4\% which is greater than might be expected. To 
see the effect of these data on the fit, a run was made with 
them left out. The result is shown in Table \ref{minima0}.

Martin \cite{martin} comments that the Damerell \ea 
\cite{damerell} data fit poorly with the rest of the data base. 
For this reason we made a fit with this data removed. Since 
there is no {\it a priori} reason to mistrust these data they 
were used in all of the other fits. The results of this fit are 
shown in Table \ref{minima0}. The normalization of the elastic 
scattering (K$^+$n) is greater than 20\% and there is some 
shift in the low-energy parameters obtained.

Two experiments have been performed \cite{edelstein,armitage} 
using the inverse charge-exchange reaction on proton targets 
with {\it K}-long beams .  It has been recognized for some time that 
there is great difficulty in controlling the normalization of 
these beams because of regeneration of {\it K}-short mesons. For this 
reason we did not use these data in the principal fit. However, 
we did include one of the sets \cite{armitage} in a run to see 
the possible effect. The results are shown in Table 
\ref{minima0}. Table \ref{k0pnorms} gives the normalization 
factors which result from the fit. The normalization errors in 
the fit were taken as 0.1 through 0.95 GeV/c and 0.2 above that 
as suggested in the experimental paper \cite{armitage}.

\begin{table}
$$\begin{array}{|c|c|}
\hline
P_{Lab}\ {\rm (GeV/c)}&{\rm Normalization\ Factor}\\
\hline
0.65&0.769\\
\hline
0.75&1.025\\
\hline
0.85&0.684\\
\hline
0.95&0.667\\
\hline
1.05&0.695\\
\hline
1.15&0.881\\
\hline
1.25&1.082\\
\hline
1.35&1.210\\
\hline
1.45&1.356\\
\hline
\end{array}$$
\caption{Normalization factors for the data of Armitage \ea 
\cite{armitage}.}\label{k0pnorms}
\end{table}

\section{Summary and Conclusions}

We have presented an easy-to-use parameterization of the {\it K}$^+$ 
Nucleon amplitudes. As mentioned in the introduction, there is
a need for reliable amplitudes for several purposes.

Our representation for s- and p-waves is equivalent to the 
effective-range expansion which, for s-waves, reads

\eq k\cot\delta=1/a+\h r_0k^2\dots \qe 
This form was derived originally for a potential interaction but 
was shown to be valid for an effective field theory 
\cite{bethe,bl}. More recently it has been shown to arise from 
renormalization group calculations \cite{birse}. Since one would 
tend to believe that this is an appropriate expansion it would seem 
to be ill advised to set the scattering length to zero.

We now discuss several points which are particularly 
interesting for their physics potential.

1) The ratio of the scattering length for {\it T}=1 to that of {\it T}=0 
   may be an interesting quantity. For example, in the work of 
   Barnes and Swanson \cite{barnes} this ratio depends only on 
   the ratio of the sizes of the kaon and nucleon and the 
   strange quark mass, being independent of the absolute 
   size of the hadronic systems. While this is only 
   approximately true in their work (i.e. only in Born 
   approximation), it indicates that this may be a quantity 
   which is sensitive to only a restricted set of physical 
   parameters. Our value for this ratio is $2.9\pm 0.3$. 

2) The reaction cross section in the {\it T}=0 state may provide an 
   interesting piece of data for the calculation of pion 
   production. Here the usual dominant mechanism ($\Delta$ 
   production) is isospin forbidden so that other mechanisms 
   will be more apparent. Calculations of the type of Oset and 
   Vicente Vacas \cite{oset} might be interesting for the {\it T}=0 
   channel.

3) The simple form of the $\ell>0$ phase shifts for the T=0 
   amplitude is remarkable. To date, no theoretical model has 
   been able to reproduce this feature althugh the work of 
   B\"{u}ttgen \ea \cite{wyborny} was able to get a moderately 
   good representation of the data with some degree of 
   phenomenology. 

4) The presence of a narrow pentaquark state would facilitate 
   the understanding of pion production in the {\it T}=1 channel. In 
   fact, the best way to look for a narrow resonance may be to 
   produce it and look for a sudden change in the inelastic 
   cross section. This was the way in which the existence of 
   the {\it J}/$\Psi$ was first indicated. If this is indeed the 
   explanation of the rapid rise in pion production and if the 
   final $\Theta-\pi$ state has relative angular momentum zero, 
   then the spin-parity of the $\Theta$ must be {\it D}5/2+. Thus, 
   the resonance, as seen directly in the {\it T}=0 channel, would be 
   in the {\it F}5/2 partial wave. We have seen that this partial 
   wave is very attractive with a large angular momentum 
   barrier so that a narrow ``molecular'' state is possible.

We thank Jean-Pierre Dedonder for comments after a careful 
reading of the manuscript.

This work was supported by the National Science Foundation under
contract PHY-0099729.

\end{document}